%%%%%%%%%%%%%%%%%%%%%%%%%%%%%%%%%%%%%%%%%%%%%%%%%%%%%%%%%%%%%%%%%%%%%
%
%
%                                      Andrea PASQUINUCCI, 1988
%              PANDA.TEX               S.I.S.S.A., Trieste, Italy
%                                      (Revised 1991, Princeton, USA)
%
%--------------------------------------------------------------------
%
%    These are TEX macros. They work with PLAIN TEX (the basis
%    version of TEX). The only problem can be with the double-page
%    format since it depends on the type of software and laserwriter
%    you use to print, so I cannot guarantee that the double-page
%    format will work properly. Double-page MUST be printed in
%    LANDSCAPE orientation. (You shouldn't have troubles with fonts;
%    if you do, please let me know.)
%
%--------------------------------------------------------------------
%
%                     INTERACTIVE SECTION
%
%--------------------------------------------------------------------
%
\def\standardrisposta{s }\def\reducedrisposta{r }
\def\mplarisposta{mpla }\def\zerorisposta{z }
\def\doublerisposta{d }\def\cartarisposta{e }\def\amsrisposta{y }
\newcount\ingrandimento \newcount\sinnota \newcount\dimnota
\newcount\unoduecol \newdimen\collhsize \newdimen\tothsize
\newdimen\fullhsize \newcount\controllorisposta \sinnota=1
\newskip\infralinea  \global\controllorisposta=0
\immediate\write16 { ********  Welcome to PANDA macros (Plain TeX,
AP, 1991) ******** }
%\immediate\write16 { You'll have to answer a few questions in
%lowercase.}
%\message{>  Do you want it in double-page (d), reduced (r)
%or standard format (s) ? }\read-1 to\risposta
%
%\message{>  Do you want it in USA A4 (u) or EUROPEAN A4
%(e) paper size ? }\read-1 to\srisposta
%
%\message{>  Do you have AMSFonts 2.0 (math) fonts (y/n) ? }
%\read-1 to\arisposta
%
%--------------------------------------------------------------------
%
%             END INTERACTIVE SECTION - PAGE FORMATTING
%
%--------------------------------------------------------------------
%       The following parameters define defaults to the interactive
%       session.  At the moment I have set EUROPEAN and MATH FONTS
%
\def\risposta{s } 
\def\srisposta{e } 
\def\arisposta{y }
\ifx\risposta\standardrisposta \ingrandimento=1200
\message {>> This will come out UNREDUCED << }
\dimnota=2 \unoduecol=1 \global\controllorisposta=1 \fi
\ifx\risposta\reducedrisposta \ingrandimento=1095 \dimnota=1
\unoduecol=1  \global\controllorisposta=1
\message {>> This will come out REDUCED << } \fi
\ifx\risposta\doublerisposta \ingrandimento=1000 \dimnota=2
\unoduecol=2

\message {>> You must print this in
LANDSCAPE orientation << } \global\controllorisposta=1 \fi
\ifx\risposta\mplarisposta \ingrandimento=1000 \dimnota=1
\message {>> Mod. Phys. Lett. A format << }
\unoduecol=1 \global\controllorisposta=1 \fi
\ifx\risposta\zerorisposta \ingrandimento=1000 \dimnota=2
\message {>> Zero Magnification format << }
\unoduecol=1 \global\controllorisposta=1 \fi
\ifnum\controllorisposta=0  \ingrandimento=1200
\message {>>> ERROR IN INPUT, I ASSUME STANDARD
UNREDUCED FORMAT <<< }  \dimnota=2 \unoduecol=1 \fi
\magnification=\ingrandimento
%
%--------------------------------------------------------------------
%
%                        PARAMETERS SETTING
%
%  You can modify these parameters at your will (and resposability)
%--------------------------------------------------------------------
%
\newdimen\eucolumnsize \newdimen\eudoublehsize \newdimen\eudoublevsize
\newdimen\uscolumnsize \newdimen\usdoublehsize \newdimen\usdoublevsize
\newdimen\eusinglehsize \newdimen\eusinglevsize \newdimen\ussinglehsize
\newskip\standardbaselineskip \newdimen\ussinglevsize
\newskip\reducedbaselineskip \newskip\doublebaselineskip
\eucolumnsize=12.0truecm    % column h-size for european doublepage
                            % (12.0treucm default)
\eudoublehsize=25.5truecm   % sheet h-size for european duoblepage
                            % (25.5treucm default)
\eudoublevsize=6.7truein    % sheet v-size for european doublepage
                            % (6.5treuin default  or 17truecm?)
\uscolumnsize=4.4truein     % column h-size for american doublepage
                            % (4.4treuin default)
\usdoublehsize=9.4truein    % sheet h-size for american duoblepage
                            % (9.4treuin default)
\usdoublevsize=6.8truein    % sheet v-size for american doublepage
                            % (6.8treuin default)
\eusinglehsize=6.5truein    % sheet h-size for european singlepage
                            % (6.5truein default)
\eusinglevsize=24truecm     % sheet v-size for european singlepage
                            % (24truecm default)
\ussinglehsize=6.5truein    % sheet h-size for american singlepage
                            % (6.5truein default)
\ussinglevsize=8.9truein    % sheet v-size for american singlepage
                            % (8.9truein default)
\standardbaselineskip=16pt plus.2pt  % baselineskip for standard
                                     % format (16pt default)
\reducedbaselineskip=14pt plus.2pt   % baselineskip for reduced
                                     % format (14pt default)
\doublebaselineskip=12pt plus.2pt    % baselineskip for doublepage
                                     % format (12pt default)
%
%  \Portoffset and \Landoffset define the horizontal and vertical
%  offsets respectively for portrait and landscape modes. Example:
%  \def\Portoffset{\voffset=.4truein\hoffset=.125truein}
%
\def\Portoffset{}
\def\Landoffset{\voffset=-.2truein}
\ifx\risposta\mplarisposta \def\Portoffset{\hoffset=1.8truecm} \fi
%
%  \Landspec defines the \special command that sets the printer
%  to landscape mode without need to specify it directly in the
%  TeX to postscript translator (the command is site dependent).
%  Example: \def\Landspec{\special{ps: landscape}}
%
\def\Landspec{}
\tolerance=10000
\parskip=0pt plus2pt  \leftskip=0pt \rightskip=0pt
%
%   Do not modify anything of what follows
%                       (unless you know what you are doing!)
%----------------------------------------------------------------------
%
\ifx\risposta\standardrisposta \infralinea=\standardbaselineskip \fi
\ifx\risposta\reducedrisposta  \infralinea=\reducedbaselineskip \fi
\ifx\risposta\doublerisposta   \infralinea=\doublebaselineskip \fi
\ifx\risposta\mplarisposta     \infralinea=13pt \fi
\ifx\risposta\zerorisposta     \infralinea=12pt plus.2pt\fi
\ifnum\controllorisposta=0    \infralinea=\standardbaselineskip \fi
\ifx\risposta\doublerisposta   \Landoffset \else \Portoffset \fi
\ifx\risposta\doublerisposta \ifx\srisposta\cartarisposta
\tothsize=\eudoublehsize \collhsize=\eucolumnsize
\vsize=\eudoublevsize  \else  \tothsize=\usdoublehsize
\collhsize=\uscolumnsize \vsize=\usdoublevsize \fi \else
\ifx\srisposta\cartarisposta \tothsize=\eusinglehsize
\vsize=\eusinglevsize \else  \tothsize=\ussinglehsize
\vsize=\ussinglevsize \fi \collhsize=4.4truein \fi
\ifx\risposta\mplarisposta \tothsize=5.0truein
\vsize=7.8truein \collhsize=4.4truein \fi
%
%--------------------------------------------------------------------
%
%                            FONTS
%
%--------------------------------------------------------------------
%
\newcount\contaeuler \newcount\contacyrill \newcount\contaams
\font\ninerm=cmr9  \font\eightrm=cmr8  \font\sixrm=cmr6
\font\ninei=cmmi9  \font\eighti=cmmi8  \font\sixi=cmmi6
\font\ninesy=cmsy9  \font\eightsy=cmsy8  \font\sixsy=cmsy6
\font\ninebf=cmbx9  \font\eightbf=cmbx8  \font\sixbf=cmbx6
\font\ninett=cmtt9  \font\eighttt=cmtt8  \font\nineit=cmti9
\font\eightit=cmti8 \font\ninesl=cmsl9  \font\eightsl=cmsl8
\skewchar\ninei='177 \skewchar\eighti='177 \skewchar\sixi='177
\skewchar\ninesy='60 \skewchar\eightsy='60 \skewchar\sixsy='60
\hyphenchar\ninett=-1 \hyphenchar\eighttt=-1 \hyphenchar\tentt=-1
\def\bfmath{\cmmib}                 % math italic bold \bfmath
\font\tencmmib=cmmib10  \newfam\cmmibfam  \skewchar\tencmmib='177
                  % math bold (cal) symbols
\font\tencmbsy=cmbsy10  \newfam\cmbsyfam  \skewchar\tencmbsy='60
\def\scaps{\cmcsc}                 % small caps (uppercase)
\font\tencmcsc=cmcsc10  \newfam\cmcscfam
\ifnum\ingrandimento=1095

\font\capsone=cmcsc10 at 10.95pt \font\capstwo=cmcsc10 at 13.145pt

\else

\font\capsone=cmcsc10 at 12pt \font\capstwo=cmcsc10 at 14.4pt
\fi

\def\ttaarr{\bf}		% chapter titles' font
\def\ppaarr{\sl}		% section titles' font

%
     % inch-high caps (enormous)
%
%   AMS fonts (this works only if you have at least the 2.0
%              version of AMSFonts, otherwise say no)
%
\newfam\eufmfam \newfam\msamfam \newfam\msbmfam \newfam\eufbfam
\def\Loadeulerfonts{\global\contaeuler=1 \ifx\arisposta\amsrisposta
\font\teneufm=eufm10              %  \eufm   Gothic (or Euler)
\font\eighteufm=eufm8 \font\nineeufm=eufm9 \font\sixeufm=eufm6
\font\seveneufm=eufm7  \font\fiveeufm=eufm5
\font\teneufb=eufb10              %  \eufb   Bold Gothic (or Euler)
\font\eighteufb=eufb8 \font\nineeufb=eufb9 \font\sixeufb=eufb6
\font\seveneufb=eufb7  \font\fiveeufb=eufb5
\font\teneurm=eurm10              %  \eurm   Roman Gothic (or Euler)
\font\eighteurm=eurm8 \font\nineeurm=eurm9
\font\teneurb=eurb10              %  \eurb   Roman Bold Gothic
\font\eighteurb=eurb8 \font\nineeurb=eurb9
\font\teneusm=eusm10              %  \eusm   Slanted Capital Gothic
\font\eighteusm=eusm8 \font\nineeusm=eusm9
\font\teneusb=eusb10              %\eusb Slanted Capital Bold Gothic
\font\eighteusb=eusb8 \font\nineeusb=eusb9
\else \def\eufm{\tt} \def\eufb{\tt} \def\eurm{\tt} \def\eurb{\tt}
\def\eusm{\tt} \def\eusb{\tt}    \fi}
\def\loadeuler{\Loadeulerfonts\tenpoint}
\def\loadamsmath{\global\contaams=1 \ifx\arisposta\amsrisposta
\font\tenmsam=msam10 \font\ninemsam=msam9 \font\eightmsam=msam8
\font\sevenmsam=msam7 \font\sixmsam=msam6 \font\fivemsam=msam5
\font\tenmsbm=msbm10 \font\ninemsbm=msbm9 \font\eightmsbm=msbm8
\font\sevenmsbm=msbm7 \font\sixmsbm=msbm6 \font\fivemsbm=msbm5
\else \def\msbm{\bf} \fi \def\Bbb{\msbm} \def\symbl{\msam} \tenpoint}
\def\loadcyrill{\global\contacyrill=1 \ifx\arisposta\amsrisposta
\font\tenwncyr=wncyr10 \font\ninewncyr=wncyr9 \font\eightwncyr=wncyr8
\font\tenwncyb=wncyr10 \font\ninewncyb=wncyr9 \font\eightwncyb=wncyr8
\font\tenwncyi=wncyr10 \font\ninewncyi=wncyr9 \font\eightwncyi=wncyr8
\else \def\cyrill{\sl} \def\cyrilb{\sl} \def\cyrili{\sl} \fi\tenpoint}
\ifx\arisposta\amsrisposta
\font\sevenex=cmex7               %  reduced math symbols
\font\eightex=cmex8  \font\nineex=cmex9
\font\ninecmmib=cmmib9   \font\eightcmmib=cmmib8
\font\sevencmmib=cmmib7 \font\sixcmmib=cmmib6
\font\fivecmmib=cmmib5   \skewchar\ninecmmib='177
\skewchar\eightcmmib='177  \skewchar\sevencmmib='177
\skewchar\sixcmmib='177   \skewchar\fivecmmib='177
\font\ninecmbsy=cmbsy9    \font\eightcmbsy=cmbsy8
\font\sevencmbsy=cmbsy7  \font\sixcmbsy=cmbsy6
\font\fivecmbsy=cmbsy5   \skewchar\ninecmbsy='60
\skewchar\eightcmbsy='60  \skewchar\sevencmbsy='60
\skewchar\sixcmbsy='60    \skewchar\fivecmbsy='60
\font\ninecmcsc=cmcsc9    \font\eightcmcsc=cmcsc8     \else
\def\cmmib{\fam\cmmibfam\tencmmib}\textfont\cmmibfam=\tencmmib
\scriptfont\cmmibfam=\tencmmib \scriptscriptfont\cmmibfam=\tencmmib
\def\cmbsy{\fam\cmbsyfam\tencmbsy} \textfont\cmbsyfam=\tencmbsy
\scriptfont\cmbsyfam=\tencmbsy \scriptscriptfont\cmbsyfam=\tencmbsy
\scriptfont\cmcscfam=\tencmcsc \scriptscriptfont\cmcscfam=\tencmcsc
\def\cmcsc{\fam\cmcscfam\tencmcsc} \textfont\cmcscfam=\tencmcsc \fi
\catcode`@=11
\newskip\ttglue
\gdef\tenpoint{\def\rm{\fam0\tenrm}
  \textfont0=\tenrm \scriptfont0=\sevenrm \scriptscriptfont0=\fiverm
  \textfont1=\teni \scriptfont1=\seveni \scriptscriptfont1=\fivei
  \textfont2=\tensy \scriptfont2=\sevensy \scriptscriptfont2=\fivesy
  \textfont3=\tenex \scriptfont3=\tenex \scriptscriptfont3=\tenex
  \def\mcal{\fam2 \tensy}  \def\mmit{\fam1 \teni}
  \textfont\itfam=\tenit \def\it{\fam\itfam\tenit}
  \textfont\slfam=\tensl \def\sl{\fam\slfam\tensl}
  \textfont\ttfam=\tentt \scriptfont\ttfam=\eighttt
  \scriptscriptfont\ttfam=\eighttt  \def\tt{\fam\ttfam\tentt}
  \textfont\bffam=\tenbf \scriptfont\bffam=\sevenbf
  \scriptscriptfont\bffam=\fivebf \def\bf{\fam\bffam\tenbf}
     \ifx\arisposta\amsrisposta    \ifnum\contaeuler=1
  \textfont\eufmfam=\teneufm \scriptfont\eufmfam=\seveneufm
  \scriptscriptfont\eufmfam=\fiveeufm \def\eufm{\fam\eufmfam\teneufm}
  \textfont\eufbfam=\teneufb \scriptfont\eufbfam=\seveneufb
  \scriptscriptfont\eufbfam=\fiveeufb \def\eufb{\fam\eufbfam\teneufb}
  \def\eurm{\teneurm} \def\eurb{\teneurb} \def\eusm{\teneusm}
  \def\eusb{\teneusb}    \fi    \ifnum\contaams=1
  \textfont\msamfam=\tenmsam \scriptfont\msamfam=\sevenmsam
  \scriptscriptfont\msamfam=\fivemsam \def\msam{\fam\msamfam\tenmsam}
  \textfont\msbmfam=\tenmsbm \scriptfont\msbmfam=\sevenmsbm
  \scriptscriptfont\msbmfam=\fivemsbm \def\msbm{\fam\msbmfam\tenmsbm}
     \fi      \ifnum\contacyrill=1     \def\cyrill{\tenwncyr}
  \def\cyrilb{\tenwncyb}  \def\cyrili{\tenwncyi}         \fi
  \textfont3=\tenex \scriptfont3=\sevenex \scriptscriptfont3=\sevenex
  \def\cmmib{\fam\cmmibfam\tencmmib} \scriptfont\cmmibfam=\sevencmmib
  \textfont\cmmibfam=\tencmmib  \scriptscriptfont\cmmibfam=\fivecmmib
  \def\cmbsy{\fam\cmbsyfam\tencmbsy} \scriptfont\cmbsyfam=\sevencmbsy
  \textfont\cmbsyfam=\tencmbsy  \scriptscriptfont\cmbsyfam=\fivecmbsy
  \def\cmcsc{\fam\cmcscfam\tencmcsc} \scriptfont\cmcscfam=\eightcmcsc
  \textfont\cmcscfam=\tencmcsc \scriptscriptfont\cmcscfam=\eightcmcsc
     \fi            \tt \ttglue=.5em plus.25em minus.15em
  \normalbaselineskip=12pt
  \setbox\strutbox=\hbox{\vrule height8.5pt depth3.5pt width0pt}
  \let\sc=\eightrm \let\big=\tenbig   \normalbaselines
  \baselineskip=\infralinea  \rm}
\gdef\ninepoint{\def\rm{\fam0\ninerm}
  \textfont0=\ninerm \scriptfont0=\sixrm \scriptscriptfont0=\fiverm
  \textfont1=\ninei \scriptfont1=\sixi \scriptscriptfont1=\fivei
  \textfont2=\ninesy \scriptfont2=\sixsy \scriptscriptfont2=\fivesy
  \textfont3=\tenex \scriptfont3=\tenex \scriptscriptfont3=\tenex
  \def\mcal{\fam2 \ninesy}  \def\mmit{\fam1 \ninei}
  \textfont\itfam=\nineit \def\it{\fam\itfam\nineit}
  \textfont\slfam=\ninesl \def\sl{\fam\slfam\ninesl}
  \textfont\ttfam=\ninett \scriptfont\ttfam=\eighttt
  \scriptscriptfont\ttfam=\eighttt \def\tt{\fam\ttfam\ninett}
  \textfont\bffam=\ninebf \scriptfont\bffam=\sixbf
  \scriptscriptfont\bffam=\fivebf \def\bf{\fam\bffam\ninebf}
     \ifx\arisposta\amsrisposta  \ifnum\contaeuler=1
  \textfont\eufmfam=\nineeufm \scriptfont\eufmfam=\sixeufm
  \scriptscriptfont\eufmfam=\fiveeufm \def\eufm{\fam\eufmfam\nineeufm}
  \textfont\eufbfam=\nineeufb \scriptfont\eufbfam=\sixeufb
  \scriptscriptfont\eufbfam=\fiveeufb \def\eufb{\fam\eufbfam\nineeufb}
  \def\eurm{\nineeurm} \def\eurb{\nineeurb} \def\eusm{\nineeusm}
  \def\eusb{\nineeusb}     \fi   \ifnum\contaams=1
  \textfont\msamfam=\ninemsam \scriptfont\msamfam=\sixmsam
  \scriptscriptfont\msamfam=\fivemsam \def\msam{\fam\msamfam\ninemsam}
  \textfont\msbmfam=\ninemsbm \scriptfont\msbmfam=\sixmsbm
  \scriptscriptfont\msbmfam=\fivemsbm \def\msbm{\fam\msbmfam\ninemsbm}
     \fi       \ifnum\contacyrill=1     \def\cyrill{\ninewncyr}
  \def\cyrilb{\ninewncyb}  \def\cyrili{\ninewncyi}         \fi
  \textfont3=\nineex \scriptfont3=\sevenex \scriptscriptfont3=\sevenex
  \def\cmmib{\fam\cmmibfam\ninecmmib}  \textfont\cmmibfam=\ninecmmib
  \scriptfont\cmmibfam=\sixcmmib \scriptscriptfont\cmmibfam=\fivecmmib
  \def\cmbsy{\fam\cmbsyfam\ninecmbsy}  \textfont\cmbsyfam=\ninecmbsy
  \scriptfont\cmbsyfam=\sixcmbsy \scriptscriptfont\cmbsyfam=\fivecmbsy
  \def\cmcsc{\fam\cmcscfam\ninecmcsc} \scriptfont\cmcscfam=\eightcmcsc
  \textfont\cmcscfam=\ninecmcsc \scriptscriptfont\cmcscfam=\eightcmcsc
     \fi            \tt \ttglue=.5em plus.25em minus.15em
  \normalbaselineskip=11pt
  \setbox\strutbox=\hbox{\vrule height8pt depth3pt width0pt}
  \let\sc=\sevenrm \let\big=\ninebig \normalbaselines\rm}
\gdef\eightpoint{\def\rm{\fam0\eightrm}
  \textfont0=\eightrm \scriptfont0=\sixrm \scriptscriptfont0=\fiverm
  \textfont1=\eighti \scriptfont1=\sixi \scriptscriptfont1=\fivei
  \textfont2=\eightsy \scriptfont2=\sixsy \scriptscriptfont2=\fivesy
  \textfont3=\tenex \scriptfont3=\tenex \scriptscriptfont3=\tenex
  \def\mcal{\fam2 \eightsy}  \def\mmit{\fam1 \eighti}
  \textfont\itfam=\eightit \def\it{\fam\itfam\eightit}
  \textfont\slfam=\eightsl \def\sl{\fam\slfam\eightsl}
  \textfont\ttfam=\eighttt \scriptfont\ttfam=\eighttt
  \scriptscriptfont\ttfam=\eighttt \def\tt{\fam\ttfam\eighttt}
  \textfont\bffam=\eightbf \scriptfont\bffam=\sixbf
  \scriptscriptfont\bffam=\fivebf \def\bf{\fam\bffam\eightbf}
     \ifx\arisposta\amsrisposta   \ifnum\contaeuler=1
  \textfont\eufmfam=\eighteufm \scriptfont\eufmfam=\sixeufm
  \scriptscriptfont\eufmfam=\fiveeufm \def\eufm{\fam\eufmfam\eighteufm}
  \textfont\eufbfam=\eighteufb \scriptfont\eufbfam=\sixeufb
  \scriptscriptfont\eufbfam=\fiveeufb \def\eufb{\fam\eufbfam\eighteufb}
  \def\eurm{\eighteurm} \def\eurb{\eighteurb} \def\eusm{\eighteusm}
  \def\eusb{\eighteusb}       \fi    \ifnum\contaams=1
  \textfont\msamfam=\eightmsam \scriptfont\msamfam=\sixmsam
  \scriptscriptfont\msamfam=\fivemsam \def\msam{\fam\msamfam\eightmsam}
  \textfont\msbmfam=\eightmsbm \scriptfont\msbmfam=\sixmsbm
  \scriptscriptfont\msbmfam=\fivemsbm \def\msbm{\fam\msbmfam\eightmsbm}
     \fi       \ifnum\contacyrill=1     \def\cyrill{\eightwncyr}
  \def\cyrilb{\eightwncyb}  \def\cyrili{\eightwncyi}         \fi
  \textfont3=\eightex \scriptfont3=\sevenex \scriptscriptfont3=\sevenex
  \def\cmmib{\fam\cmmibfam\eightcmmib}  \textfont\cmmibfam=\eightcmmib
  \scriptfont\cmmibfam=\sixcmmib \scriptscriptfont\cmmibfam=\fivecmmib
  \def\cmbsy{\fam\cmbsyfam\eightcmbsy}  \textfont\cmbsyfam=\eightcmbsy
  \scriptfont\cmbsyfam=\sixcmbsy \scriptscriptfont\cmbsyfam=\fivecmbsy
  \def\cmcsc{\fam\cmcscfam\eightcmcsc} \scriptfont\cmcscfam=\eightcmcsc
  \textfont\cmcscfam=\eightcmcsc \scriptscriptfont\cmcscfam=\eightcmcsc
     \fi             \tt \ttglue=.5em plus.25em minus.15em
  \normalbaselineskip=9pt
  \setbox\strutbox=\hbox{\vrule height7pt depth2pt width0pt}
  \let\sc=\sixrm \let\big=\eightbig \normalbaselines\rm }
\gdef\tenbig#1{{\hbox{$\left#1\vbox to8.5pt{}\right.\n@space$}}}
\gdef\ninebig#1{{\hbox{$\textfont0=\tenrm\textfont2=\tensy
   \left#1\vbox to7.25pt{}\right.\n@space$}}}
\gdef\eightbig#1{{\hbox{$\textfont0=\ninerm\textfont2=\ninesy
   \left#1\vbox to6.5pt{}\right.\n@space$}}}
 %for 10-pt math in 9-pt territory
\def\alternativefont#1#2{\ifx\arisposta\amsrisposta \relax \else
\xdef#1{#2} \fi}
\global\contaeuler=0 \global\contacyrill=0 \global\contaams=0
%
%--------------------------------------------------------------------
%
%                            MACROS
%
%--------------------------------------------------------------------
%
\newbox\fotlinebb \newbox\hedlinebb \newbox\leftcolumn
\gdef\makeheadline{\vbox to 0pt{\vskip-22.5pt
     \fullline{\vbox to8.5pt{}\the\headline}\vss}\nointerlineskip}
\gdef\makehedlinebb{\vbox to 0pt{\vskip-22.5pt
     \fullline{\vbox to8.5pt{}\copy\hedlinebb\hfil
     \line{\hfill\the\headline\hfill}}\vss} \nointerlineskip}
\gdef\makefootline{\baselineskip=24pt \fullline{\the\footline}}
\gdef\makefotlinebb{\baselineskip=24pt
    \fullline{\copy\fotlinebb\hfil\line{\hfill\the\footline\hfill}}}
\gdef\doubleformat{\shipout\vbox{\Landspec\makehedlinebb
     \fullline{\box\leftcolumn\hfil\columnbox}\makefotlinebb}
     \advancepageno}
\gdef\columnbox{\leftline{\pagebody}}
\gdef\line#1{\hbox to\hsize{\hskip\leftskip#1\hskip\rightskip}}
\gdef\fullline#1{\hbox to\fullhsize{\hskip\leftskip{#1}%
\hskip\rightskip}}
\gdef\footnote#1{\let\@sf=\empty
         \ifhmode\edef\#sf{\spacefactor=\the\spacefactor}\/\fi
         #1\@sf\vfootnote{#1}}
\gdef\vfootnote#1{\insert\footins\bgroup
         \ifnum\dimnota=1  \eightpoint\fi
         \ifnum\dimnota=2  \ninepoint\fi
         \ifnum\dimnota=0  \tenpoint\fi
         \interlinepenalty=\interfootnotelinepenalty
         \splittopskip=\ht\strutbox
         \splitmaxdepth=\dp\strutbox \floatingpenalty=20000
         \leftskip=\oldssposta \rightskip=\olddsposta
         \spaceskip=0pt \xspaceskip=0pt
         \ifnum\sinnota=0   \textindent{#1}\fi
         \ifnum\sinnota=1   \item{#1}\fi
         \footstrut\futurelet\next\fo@t}
\gdef\fo@t{\ifcat\bgroup\noexpand\next \let\next\f@@t
             \else\let\next\f@t\fi \next}
\gdef\f@@t{\bgroup\aftergroup\@foot\let\next}
\gdef\f@t#1{#1\@foot} \gdef\@foot{\strut\egroup}
\gdef\footstrut{\vbox to\splittopskip{}}
\skip\footins=\bigskipamount
\count\footins=1000  \dimen\footins=8in
\catcode`@=12
\tenpoint
\ifnum\unoduecol=1 \hsize=\tothsize   \fullhsize=\tothsize \fi
\ifnum\unoduecol=2 \hsize=\collhsize  \fullhsize=\tothsize \fi
\global\let\lrcol=L      \ifnum\unoduecol=1
\output{\plainoutput{\ifnum\tipbnota=2 \clearnmbnota\fi}} \fi
\ifnum\unoduecol=2 \output{\if L\lrcol
     \global\setbox\leftcolumn=\columnbox
     \global\setbox\fotlinebb=\line{\hfill\the\footline\hfill}
     \global\setbox\hedlinebb=\line{\hfill\the\headline\hfill}
     \advancepageno  \global\let\lrcol=R
     \else  \doubleformat \global\let\lrcol=L \fi
     \ifnum\outputpenalty>-20000 \else\dosupereject\fi
     \ifnum\tipbnota=2\clearnmbnota\fi }\fi
\def\ifdoublepage{\ifnum\unoduecol=2 }
\gdef\yespagenumbers{\footline={\hss\tenrm\folio\hss}}
\gdef\ciao{ \ifnum\fdefcontre=1 \endfdef\fi
     \par\vfill\supereject \ifnum\unoduecol=2
     \if R\lrcol  \headline={}\nopagenumbers\null\vfill\eject
     \fi\fi \end}

\newskip\olddsposta \newskip\oldssposta
\global\oldssposta=\leftskip \global\olddsposta=\rightskip

\def\filldots{\leaders\hbox to 1em{\hss.\hss}\hfill}
\def\inquadrb#1 {\vbox {\hrule  \hbox{\vrule \vbox {\vskip .2cm
    \hbox {\ #1\ } \vskip .2cm } \vrule  }  \hrule} }
 \def\newline{\hfil\break}
\def\jump{\vskip\baselineskip} \newskip\iinnffrr
\def\sjump{\iinnffrr=\baselineskip
          \divide\iinnffrr by 2 \vskip\iinnffrr}
\def\bjump{\vskip\baselineskip \vskip\baselineskip}
\newcount\nmbnota  \def\clearnmbnota{\global\nmbnota=0}
\newcount\tipbnota \def\letterfootnote{\global\tipbnota=1}

\def\note#1{\global\advance\nmbnota by 1 \ifnum\tipbnota=1
    \footnote{$^{\rm\nttlett}$}{#1} \else {\ifnum\tipbnota=2
    \footnote{$^{\nttsymb}$}{#1}
    \else\footnote{$^{\the\nmbnota}$}{#1}\fi}\fi}
\def\nttlett{\ifcase\nmbnota \or a\or b\or c\or d\or e\or f\or
g\or h\or i\or j\or k\or l\or m\or n\or o\or p\or q\or r\or
s\or t\or u\or v\or w\or y\or x\or z\fi}
\def\nttsymb{\ifcase\nmbnota \or\dag\or\sharp\or\ddag\or\star\or
\natural\or\flat\or\clubsuit\or\diamondsuit\or\heartsuit
\or\spadesuit\fi}   \clearnmbnota
\def\numberfootnote{\global\tipbnota=0} \numberfootnote
\def\setnote#1{\expandafter\xdef\csname#1\endcsname{
\ifnum\tipbnota=1 {\rm\nttlett} \else {\ifnum\tipbnota=2
{\nttsymb} \else \the\nmbnota\fi}\fi} }
\newcount\nbmfig  \def\clearnbmfig{\global\nbmfig=0}
\gdef\figure{\global\advance\nbmfig by 1
      {\rm fig. \the\nbmfig}}   \clearnbmfig
\def\setfig#1{\expandafter\xdef\csname#1\endcsname{fig. \the\nbmfig}}
 \def\endformula{\eqno\numero $$}
 \def\efr{\endformula}
\newcount\frmcount \def\clearfrmcount{\global\frmcount=0}
\def\numero{\global\advance\frmcount by 1   \ifnum\indappcount=0
  {\ifnum\cpcount <1 {\hbox{\rm (\the\frmcount )}}  \else
  {\hbox{\rm (\the\cpcount .\the\frmcount )}} \fi}  \else
  {\hbox{\rm (\applett .\the\frmcount )}} \fi}
\def\nameformula#1{\global\advance\frmcount by 1%
\ifnum\draftnum=0  {\ifnum\indappcount=0%
{\ifnum\cpcount<1\xdef\spzzttrra{(\the\frmcount )}%
\else\xdef\spzzttrra{(\the\cpcount .\the\frmcount )}\fi}%
\else\xdef\spzzttrra{(\applett .\the\frmcount )}\fi}%
\else\xdef\spzzttrra{(#1)}\fi%
\expandafter\xdef\csname#1\endcsname{\spzzttrra}
\eqno \hbox{\rm\spzzttrra} $$}
\def\nfr{\nameformula}    \def\numali{\numero}
\def\nameali#1{\global\advance\frmcount by 1%
\ifnum\draftnum=0  {\ifnum\indappcount=0%
{\ifnum\cpcount<1\xdef\spzzttrra{(\the\frmcount )}%
\else\xdef\spzzttrra{(\the\cpcount .\the\frmcount )}\fi}%
\else\xdef\spzzttrra{(\applett .\the\frmcount )}\fi}%
\else\xdef\spzzttrra{(#1)}\fi%
\expandafter\xdef\csname#1\endcsname{\spzzttrra}
  \hbox{\rm\spzzttrra} }      \clearfrmcount
\newcount\cpcount \def\clearcpcount{\global\cpcount=0}
\newcount\subcpcount \def\clearsubcpcount{\global\subcpcount=0}
\newcount\appcount \def\clearappcount{\global\appcount=0}
\newcount\indappcount \def\clearindappcount{\indappcount=0}
\newcount\sottoparcount 

\def\applett{\ifcase\appcount  \or {A}\or {B}\or {C}\or
{D}\or {E}\or {F}\or {G}\or {H}\or {I}\or {J}\or {K}\or {L}\or
{M}\or {N}\or {O}\or {P}\or {Q}\or {R}\or {S}\or {T}\or {U}\or
{V}\or {W}\or {X}\or {Y}\or {Z}\fi    \ifnum\appcount<0
\immediate\write16 {Panda ERROR - Appendix: counter "appcount"
out of range}\fi  \ifnum\appcount>26  \immediate\write16 {Panda
ERROR - Appendix: counter "appcount" out of range}\fi}
\clearappcount  \clearindappcount \newcount\connttrre
\def\clearconnttrre{\global\connttrre=0} \newcount\countref
\def\clearcountref{\global\countref=0} \clearcountref
\def\chapter#1{\global\advance\cpcount by 1 \clearfrmcount
                 \goodbreak\null\vbox{\jump\nobreak
                 \clearsubcpcount\clearindappcount
                 \itemitem{\ttaarr\the\cpcount .\qquad}{\ttaarr #1}
                 \par\nobreak\jump\sjump}\nobreak}
\def\section#1{\global\advance\subcpcount by 1 \goodbreak\null
               \vbox{\sjump\nobreak\ifnum\indappcount=0
                 {\ifnum\cpcount=0 {\itemitem{\ppaarr
               .\the\subcpcount\quad\enskip\ }{\ppaarr #1}\par} \else
                 {\itemitem{\ppaarr\the\cpcount .\the\subcpcount\quad
                  \enskip\ }{\ppaarr #1} \par}  \fi}
                \else{\itemitem{\ppaarr\applett .\the\subcpcount\quad
                 \enskip\ }{\ppaarr #1}\par}\fi\nobreak\jump}\nobreak}
\clearsubcpcount
\def\appendix#1{\global\advance\appcount by 1 \clearfrmcount
                  \goodbreak\null\vbox{\jump\nobreak
                  \global\advance\indappcount by 1 \clearsubcpcount
          \itemitem{ }{\hskip-40pt\ttaarr #1}
%                  \itemitem{\ttaarr App.\applett\ }{\ttaarr #1}
             \nobreak\jump\sjump}\nobreak}
\clearappcount \clearindappcount
\def\references{\goodbreak\null\vbox{\jump\nobreak
   \noindent{\ttaarr References} \nobreak\jump\sjump}\nobreak}
%   \itemitem{}{\ttaarr References} \nobreak\jump\sjump}\nobreak}

\clearcpcount\clearcountref

\def\setchap#1{\ifnum\indappcount=0{\ifnum\subcpcount=0%
\xdef\spzzttrra{\the\cpcount}%
\else\xdef\spzzttrra{\the\cpcount .\the\subcpcount}\fi}
\else{\ifnum\subcpcount=0 \xdef\spzzttrra{\applett}%
\else\xdef\spzzttrra{\applett .\the\subcpcount}\fi}\fi
\expandafter\xdef\csname#1\endcsname{\spzzttrra}}
\newcount\draftnum \newcount\ppora   \newcount\ppminuti
\global\ppora=\time   \global\ppminuti=\time
\global\divide\ppora by 60  \draftnum=\ppora
\multiply\draftnum by 60    \global\advance\ppminuti by -\draftnum
\def\droggi{\number\day /\number\month /\number\year\ \the\ppora
:\the\ppminuti}     \global\draftnum=0
\def\draftcomment#1{\ifnum\draftnum=0 \relax \else
{\ {\bf ***}\ #1\ {\bf ***}\ }\fi} 
%
%     Maximum number of references = 200
%     boxes 50 -> 250 reserved for references
%
\catcode`@=11
\gdef\Ref#1{\expandafter\ifx\csname @rrxx@#1\endcsname\relax%
{\global\advance\countref by 1    \ifnum\countref>200
\immediate\write16 {Panda ERROR - Ref: maximum number of references
exceeded}  \expandafter\xdef\csname @rrxx@#1\endcsname{0}\else
\expandafter\xdef\csname @rrxx@#1\endcsname{\the\countref}\fi}\fi
\ifnum\draftnum=0 \csname @rrxx@#1\endcsname \else#1\fi}
\gdef\beginref{\ifnum\draftnum=0  \gdef\Rref{\fairef}
\gdef\endref{\scriviref} \else\relax\fi
\ifx\risposta\mplarisposta \ninepoint \fi
\parskip 2pt plus.2pt \baselineskip=12pt}
\def\Reflab#1{[#1]} \gdef\Rref#1#2{\item{\Reflab{#1}}{#2}}
\gdef\endref{\relax}  \newcount\conttemp
\gdef\fairef#1#2{\expandafter\ifx\csname @rrxx@#1\endcsname\relax
{\global\conttemp=0 \immediate\write16 {Panda ERROR - Ref: reference
[#1] undefined}} \else
{\global\conttemp=\csname @rrxx@#1\endcsname } \fi
\global\advance\conttemp by 50  \global\setbox\conttemp=\hbox{#2} }
\gdef\scriviref{\clearconnttrre\conttemp=50
\loop\ifnum\connttrre<\countref \advance\conttemp by 1
\advance\connttrre by 1
\item{\Reflab{\the\connttrre}}{\unhcopy\conttemp} \repeat}
\clearcountref \clearconnttrre
\catcode`@=12
\ifx\risposta\mplarisposta \def\Reflab#1{#1.} \letterfootnote \fi

\def\slashchar#1{\setbox0=\hbox{$#1$} \dimen0=\wd0
     \setbox1=\hbox{/} \dimen1=\wd1 \ifdim\dimen0>\dimen1
      \rlap{\hbox to \dimen0{\hfil/\hfil}} #1 \else
      \rlap{\hbox to \dimen1{\hfil$#1$\hfil}} / \fi}
\ifx\oldchi\undefined \let\oldchi=\chi
  \def\cchi{{\raise 1pt\hbox{$\oldchi$}}} \let\chi=\cchi \fi

\def\frac#1#2{{\textstyle{#1 \over #2}}}

\def\half{\ifinner {\scriptstyle {1 \over 2}}\else {1 \over 2} \fi}

\def\simge{\rlap{\raise 2pt \hbox{$>$}}{\lower 2pt \hbox{$\sim$}}}
\def\simle{\rlap{\raise 2pt \hbox{$<$}}{\lower 2pt \hbox{$\sim$}}}

\def\vbig#1#2{{\vbigd@men=#2\divide\vbigd@men by 2%
\hbox{$\left#1\vbox to \vbigd@men{}\right.\n@space$}}}

%
%--------------------------------------------------------------------
%
\newcount\fdefcontre \newcount\fdefcount \newcount\indcount
\newread\filefdef  \newread\fileftmp  \newwrite\filefdef
\newwrite\fileftmp     \def\strip#1*.A {#1}
\def\futuredef#1{\beginfdef
\expandafter\ifx\csname#1\endcsname\relax%
{\immediate\write\fileftmp {#1*.A}
\immediate\write16 {Panda Warning - fdef: macro "#1" on page
\the\pageno \space undefined}
\ifnum\draftnum=0 \expandafter\xdef\csname#1\endcsname{(?)}
\else \expandafter\xdef\csname#1\endcsname{(#1)} \fi
\global\advance\fdefcount by 1}\fi   \csname#1\endcsname}

\def\beginfdef{\ifnum\fdefcontre=0
\immediate\openin\filefdef \jobname.fdef
\immediate\openout\fileftmp \jobname.ftmp
\global\fdefcontre=1  \ifeof\filefdef \immediate\write16 {Panda
WARNING - fdef: file \jobname.fdef not found, run TeX again}
\else \immediate\read\filefdef to\spzzttrra
\global\advance\fdefcount by \spzzttrra
\indcount=0      \loop\ifnum\indcount<\fdefcount
\advance\indcount by 1   \immediate\read\filefdef to\spezttrra
\immediate\read\filefdef to\sppzttrra
\edef\spzzttrra{\expandafter\strip\spezttrra}
\immediate\write\fileftmp {\spzzttrra *.A}
\expandafter\xdef\csname\spzzttrra\endcsname{\sppzttrra}
\repeat \fi \immediate\closein\filefdef \fi}
\def\endfdef{\immediate\closeout\fileftmp   \ifnum\fdefcount>0
\immediate\openin\fileftmp \jobname.ftmp
\immediate\openout\filefdef \jobname.fdef
\immediate\write\filefdef {\the\fdefcount}   \indcount=0
\loop\ifnum\indcount<\fdefcount    \advance\indcount by 1
\immediate\read\fileftmp to\spezttrra
\edef\spzzttrra{\expandafter\strip\spezttrra}
\immediate\write\filefdef{\spzzttrra *.A}
\edef\spezttrra{\string{\csname\spzzttrra\endcsname\string}}
\iwritel\filefdef{\spezttrra}
\repeat  \immediate\closein\fileftmp \immediate\closeout\filefdef
\immediate\write16 {Panda Warning - fdef: Label(s) may have changed,
re-run TeX to get them right}\fi}
\def\iwritel#1#2{\newlinechar=-1
{\newlinechar=`\ \immediate\write#1{#2}}\newlinechar=-1}
\global\fdefcontre=0 \global\fdefcount=0 \global\indcount=0
%
%--------------------------------------------------------------------
%
\null
%
%--------------------------------------------------------------------
%
%                             THE    END
%
%--------------------------------------------------------------------
%
%\input panda
%\draftmode{Quantum Integrability}
\loadamsmath
\loadeuler
\mathchardef\hbar="2D7D
\mathchardef\blamb="7315
\mathchardef\bmu="7316
\mathchardef\balpha="710B
\mathchardef\bbeta="710C
\mathchardef\bgamma="710D
\def\lab{{\bfmath\blamb}}
\def\mub{{\bfmath\bmu}}
\def\alb{{\bfmath\balpha}}
\def\beb{{\bfmath\bbeta}}
\def\gab{{\bfmath\bgamma}}
\pageno=0
\nopagenumbers{\baselineskip=12pt
\line{\hfill US-FT/44-96}
\line{\hfill SWAT/138} 
\line{\hfill\tt hep-th/9701109}
\line{\hfill January 1997}\bjump
\ifdoublepage \bjump\bjump\bjump\else\jump\vfill\fi
\centerline{\capstwo Solitonic Integrable Perturbations}
\sjump
\centerline{\capstwo of Parafermionic Theories}
\bjump\jump
\centerline{{\scaps Carlos R.~Fern\'andez-Pousa$^{\>{\rm a}}$}, 
{\scaps Manuel~V. Gallas$^{\>{\rm a}}$},} \sjump
\centerline{{\scaps Timothy J.~Hollowood$^{\>{\rm b}}$}
and {\scaps J. Luis Miramontes$^{\>{\rm a}}$}}
\jump\jump
\centerline{\sl $^{{\rm a}\>}$Departamento de F\'\i sica de Part\'\i
culas,}
\centerline{\sl Facultad de F\'\i sica,}
\centerline{\sl Universidad de Santiago,}
\centerline{\sl E-15706 Santiago de Compostela, Spain}
\jump
\centerline{\sl $^{{\rm b}\>}$Department of Physics,}
\centerline{\sl University of Wales Swansea,}
\centerline{\sl Singleton Park, Swansea SA2 8PP, U.K.}
\jump
\centerline{\tt pousa@gaes.usc.es,~ gallas@fpaxp1.usc.es} \sjump  
\centerline{{\tt t.hollowood@swansea.ac.uk},~ and~ 
{\tt miramont@fpaxp1.usc.es}} 
\bjump\bjump
\ifdoublepage
\vfill {\noindent
\line{January 1997\hfill}}
\eject\null\vfill\fi
\centerline{\capsone ABSTRACT}\jump

\noindent
The quantum integrability of a class of
massive perturbations of the parafermionic conformal field theories
associated to compact Lie groups is established by showing that they
have quantum conserved densities of scale dimension~2 and~3. These
theories are integrable for any value of a continuous vector coupling
constant, and they generalize the perturbation of the minimal
parafermionic models by their first thermal operator. The classical
equations-of-motion of these perturbed theories
are the non-abelian affine Toda equations which admit
(charged) soliton solutions whose semi-classical quantization is
expected to permit the identification of the exact $S$-matrix of the
theory. 

\vfill
\ifdoublepage \else
\noindent
\line{January 1997\hfill}\fi
\eject}
\yespagenumbers\pageno=1
\footline={\hss\tenrm-- \folio\ --\hss}

\chapter{Introduction}

This paper is concerned with establishing the 
quantum integrability of certain
perturbations of the parafermionic conformal field theories
associated to compact simple Lie groups. Classically, these
perturbations arise in the context of non-abelian affine Toda
equations, and they correspond to the ``Homogeneous sine-Gordon
theories'' of Ref.~[\Ref{MASS}]. At the  quantum level, we will show that
these theories exhibit conserved densities of (at least)  scale
dimension~2 and~3, which implies, via the usual folklore, 
the factorization of their scattering
matrices~[\Ref{THREE}] and hence their full quantum integrability.

The non-abelian affine Toda (NAAT) equations are integrable
generalizations of the sine-Gordon equations-of-motion for a bosonic
field that takes values in some, generally non-abelian, Lie
group~[\Ref{NAT}]. In~[\Ref{MASS}], we have determined the class of
NAAT equations that can be derived as the classical 
equations-of-motion of a theory whose action is real,
positive-definite, and exhibits a mass-gap. The result is that,
besides the usual abelian affine Toda equations  with real coupling
constant, the NAAT equations give rise to only two series of models, 
referred to as the Symmetric-Space Sine-Gordon theories and the
Homogeneous Sine-Gordon theories. Both are classically integrable
and, unlike the abelian affine Toda equations for real values of the
coupling constant~[\Ref{HOLL}], admit soliton
solutions. It is
important for us to emphasize that the sine-Gordon theory is the only
abelian affine Toda equation with imaginary coupling constant that is
included in this class.

In the present paper, we only consider the Homogeneous sine-Gordon
theories (HSG), which, at the classical level, are particular examples
of the deformed coset models constructed by Park in
Ref.~[\Ref{PARK1}]. The condition
that the theory has a mass-gap constrains
the coset to be of the form $G/U(1)^{\times r_g}$~[\Ref{MASS}], with
$G$ a compact simple Lie group of rank $r_g$ in which the field of the
theory takes values. At the quantum level, the coupling constant
becomes quantized and gives rise to an integer number~$k>0$, known as
the ``level''. It is in this way that 
the resulting theories are nothing other than
perturbations of the $G$-parafermion theories of 
level-$k$~[\Ref{GPARAF}]. The simplest representative of this
class is the HSG associated to the group $SU(2)$, {\it i.e.} to the
coset $SU(2)/U(1)$, whose equation-of-motion is the complex 
sine-Gordon equation. As shown by Bakas~[\Ref{BAK}], this theory
corresponds to the perturbation of the ${\Bbb Z}_k$ parafermions by
its first thermal operator, one of the three known
integrable perturbations in these models~[\Ref{PINT}]. 
However, the description of
this well-known theory within the framework of NAAT equations has
two important advantages that apply to all the other HSG theories.
Firstly, the  quantum conserved densities can be obtained through the
renormalization of the classical ones. Secondly,
since the relevant equations admit soliton solutions, the quantum
description of these theories can be elucidated by means of a
semi-classical approach, as in the sine-Gordon theory. For the HSG
associated to $SU(2)$, the semi-classical analysis leading to the
exact spectrum and $S$-matrix of the theory has been performed
in~[\Ref{TN}], and it agrees with the results of~[\Ref{PINT}]. The
study of the soliton spectrum for general HSG theories will be
presented elsewhere.    
 
Given a simple compact Lie group $G$, the construction of a HSG theory
involves the choice of two constant elements $\Lambda_+$ and
$\Lambda_-$ in some Cartan (maximal abelian) subalgebra of $g$, the
Lie algebra of $G$. These two elements fix the form of the potential
term of the Lagrangian, which is proportional to $\langle \Lambda_+ ,
h^\dagger
\Lambda_- h\rangle$, where $h$ is the HSG field taking values in $G$ 
and $\langle\>, \>\rangle$ is the Killing form of $g$. This potential
term specifies both the mass spectrum and the precise form of the
perturbation of the G-parafermionic theory. To ensure that the theory
has a mass-gap, $\Lambda_\pm$ have to be chosen such that they are
regular, which means that they cannot be orthogonal to any root of 
$g$; otherwise, their value remains arbitrary. However, different
choices of $\Lambda_\pm$ lead to different theories. 
For instance, the resulting theories are not parity invariant
unless $\Lambda_+ = h_{0}^\dagger \Lambda_- h_0$, where~$h_0$ is the
field configuration corresponding to the vacuum~[\Ref{MASS}].
Therefore, $\Lambda_\pm$ have to be considered  as continuous
%%%%%
%%%%%LUIS: I have added something about \tau here. More in the next
%%%%%section.
%%%%%
(relevant) vector coupling constants of the theory. Moreover, the role
of the kinetic term is played by a gauged WZW action associated to the
coset $G/U(1)^{\times r_g}$, whose precise form is specified by an
automorphism $\tau$ of the Cartan subalgebra of $g$ that preserves the
Killing form $\langle\>, \>\rangle$. Consequently, $\tau$~can be
(almost) any element of the orthogonal group~$O(r_g)$, and hence it
provides an additional set of continuous (marginal) coupling constants.
Remarkably we shall find that the HSG theories are quantum integrable
for any value of $\Lambda_\pm$ and $\tau$. This means that these
theories are counter-examples to the generally held belief that
integrability is a very special property which requires carefully tuned
coupling constants (apart from couplings that can be absorbed into
$\hbar$ or an overall mass scale).

The paper is organized as follows. In Section~2, we summarize
the main features of the classical HSG theories following~[\Ref{MASS}]. 
In Section~3, we show that they are
classically integrable by exhibiting the existence of an infinite
number of classical conserved charges. To be more specific, for each
positive integer scale dimension $s$, there are ${\rm rank}(G)$
linearly independent conserved densities. Explicit expressions for
$s=2$ and~3 densities are provided in the Appendix. Section~4 is the
core of the paper where quantum conserved densities of scale
dimension~2 and~3 are constructed. The HSG theories are
perturbations of a coset conformal field theory (CFT) by a relevant
primary field, and hence the existence of quantum conserved
densities can be investigated by using the method proposed by
%%%LUIS: I think that it is fair to mention the work of Delius.
%%%
Zamolodchikov in~[\Ref{ZAM}], which has already been used in the
context of abelian affine Toda theories~[\Ref{DELIUS}]. Following the
approach of Bais {\it et al.}~[\Ref{BAIS2}], the CFT associated to a
coset $G/H$ can be described in terms of the Wess-Zumino-Witten (WZW)
theory corresponding to the Lie group~$G$. Since, in our case, the
underlying ultra-violet CFT is associated to a coset of the form
$G/U(1)^{\times r_g}$, the use of this method involves an analysis of
operator products in the WZW theory associated to $G$. Our study is
restricted to the first order in perturbation theory; however, at
least for values of the level $k\geq$ the dual Coxeter number of $G$,
our results are expected to be exact~[\Ref{ZAM},\Ref{CARDY}]. Finally,
in Section~5, we present our conclusions.

\chapter{The Homogeneous sine-Gordon theories}

According to~[\Ref{MASS}], where more details can be found, the
construction of the different HSG associated to a given compact Lie
group $G$ starts with the choice of two constant elements
$\Lambda_\pm$ in $g$, the Lie algebra of $G$. The condition
that the resulting theory has a mass-gap requires $\Lambda_\pm$ 
to be semisimple and regular, which means that their centralizer in
$g$ is a Cartan subalgebra. Otherwise, the choice of $\Lambda_\pm$ is
completely free and therefore they are to be regarded as continuous
vector coupling constants of the theory. To comply with the notation 
of~[\Ref{MASS}], we will refer to the centralizer of
$\Lambda_+$ in $g$ and to the corresponding abelian group as $g_{0}^0
\simeq u(1)^{+ r_g}$ and $G_{0}^0\simeq U(1)^{\times r_g}$,
respectively, where $r_g$ is the  rank of $G$. The HSG is specified by
the action
$$
S[h, A_\pm]\>=\> {1\over \beta^2} \biggl\{ S_{\rm WZW}[h,A_\pm] 
\>-\>\int d^2 x \>V(h) \biggl\}\>.
\nfr{Act}
Here, $h$ is a bosonic field that takes values in $G$, $A_\pm$ are
(non-dymanical) abelian gauge connections taking values in $g_{0}^0$,
and $S_{\rm WZW}[h,A_\pm]$ is the gauged Wess-Zumino-Witten
action for the coset $G/G_{0}^0 \simeq G/U(1)^{\times r_g}$,  
$$
\eqalign{
S_{\rm WZW}[h & ,A_\pm]\> = \> S_{\rm WZW}[h]\> + \> {1\over 2\pi}  
\int d^2x \> \Bigl( -\langle A_+\>,\> \partial_-h\>
h^{\dagger}\rangle\cr  
& + \> \langle \tau(A_-)\>, \>
h^{\dagger}\> \partial_+h\rangle\> +\> \langle h^{\dagger}\> A_+\> 
h\>,
\>
\tau(A_-) \rangle\> -\> \langle A_+\>,\>
A_-\rangle\Bigr)\>,\cr}
\nfr{WZW}
where $x_\pm = t\pm x$ are light-cone variables in $1+1$ 
Minkowski space. The potential
$V(h)$ is
$$
V(h)\> =\> - \> {m^2 \over 2\pi}\> \langle  \Lambda_+\>,\>  
h^{\dagger}
\>
\Lambda_- \> h\rangle\>,
\nfr{Pot}
where we have denoted the Killing form on $g$ by $\langle\>
,\>\rangle$, normalized such that long roots have square length 2.
Finally, $\hbar m$ is a constant with dimensions of mass, and $\beta$ is
a coupling constant that has to be quantized if the quantum theory is
to be well defined; namely, $\hbar\>\beta^2 =1/k$, where $k$ is a
non-negative (dimensionless) integer~[\Ref{WITTEN}]. The Planck
constant is explicitly shown to exhibit that, just as in the
sine-Gordon theory, the semi-classical limit is the same as the
weak-coupling limit, and  that both are recovered when $k\rightarrow
\infty$. The action~\Act\ is invariant with respect to the abelian
gauge transformations
$$
h(x,t)\mapsto {\rm e\>}^\alpha\> h \> {\rm
e\>}^{-\>\tau(\alpha)}\>, \qquad A_\pm \mapsto A_\pm \>-
\>\partial_\pm\alpha\>,
\nfr{Gtrans}
where $\alpha=\alpha(x,t)$ takes values in $g_{0}^0
\simeq u(1)^{+ r_g}$. 

%%%
%%%LUIS: I hope that this clarifies the meaning of tau. 
%%%
It is well known that a coset~$G/H$ specifies different conformal field
theories associated to the anomaly-free embeddings of~$H$ into $G_L
\otimes G_R$, the internal chiral symmetry group of the WZW theory
corresponding to~$G$, and that all the resulting theories share the same
central charge~[\Ref{GAUGE}]. In our case, the possible anomaly-free
embeddings of~$G_{0}^0\simeq U(1)^{\times r_g}$ into $G_L \otimes G_R$
are specified by~$\tau$, which is an arbitrary automorphism of
$g_{0}^0$ that preserves the restriction of the Killing form $\langle\>
,\>\rangle$ to $g_{0}^0$. Therefore, since $g_{0}^0$ is
abelian, $\tau$ can be any element of the orthogonal group~$O(r_g)$. In 
particular, $\tau$~can always be chosen to be~$+I$ or~$-I$, which lead
to gauge transformations of vector or axial type, respectively, and
which are the only possible choices for~$G=SU(2)$. However, in order
to ensure that the resulting HSG theory has a mass-gap, $\tau$~has to be
different to the conjugation induced by the field configuration
corresponding to the vacuum~$h_{0}$, which means that $\tau(u) \not=
h_{0}^\dagger u h_0$ for any $u\in g_{0}^0$~[\Ref{MASS}].
Nevertheless, the proof of quantum integrability presented in Section~4
is based on the underlying (Kac-Moody) current-algebra structure of the
gauged WZW action specified by~$G/U(1)^{\times r_g}$, and hence it
does not make any reference to~$\tau$. Therefore,
since the different CFT's associated to the same coset are related
by means of deformations induced by marginal operators of the form~$\int
d^2x \> C_{AB} J^A(z) \> \overline{J^B}(\overline{z})$ involving the
Cartan components of the affine chiral currents~[\Ref{MARG}], the
automorphism~$\tau$ provides an additional set of continuous (marginal)
coupling constants of the HSG theories.

At the quantum level, the HSG theories can be described as a perturbed
CFT of the form
$$
S\> = \> S_{\rm CFT}\> + \>{m^2\over 2\pi\beta^2}\> \int d^2 x
\>\Phi(x,t)\>.
\nfr{QAct}
Here, $S_{\rm CFT}$ is the action of the CFT associated to the coset
$G/U(1)^{\times r_g}$ at level~$k$, which is nothing else than the
theory of level-$k$ $G$-parafermions~[\Ref{GPARAF}], whose central 
charge is
$$
C_{\rm CFT}(G,k)\> =\> {k\> {\rm dim\>}(G) \over k + h_{g}^\vee} \> 
-\> {\rm rank\>}(G)\>,
\nfr{Charge}
where $h_{g}^\vee$ is the dual Coxeter number of $G$. 

Following the work of Bais {\it et al.}~[\Ref{BAIS2}], the generators 
of the operator algebra of the coset CFT can be realized as a subset
of  the generators of the operator algebra of the underlying
Wess-Zumino-Witten model associated to $G$. Those generators include
the chiral currents $J^a(z)$ and
$\overline{J^a}(\overline{z})$~[\Ref{KZ}] that satisfy the operator
product expansion (OPE)
$$
J^a(z)\> J^b(w)\> = \> -{\hbar^2 k\over (z-w)^2}\> \delta^{ab} \> 
+\> \hbar\> f^{abc}\> {J^c(w)\over z-w} \>+ \> \cdots\>.
\nfr{OpeJ}
In this paper,  we will follow the conventions
of~[\Ref{BAIS2},\Ref{BAIS1}], and we have introduced a notation
reminiscent of euclidean space:
$z\equiv x_-$ and $\overline{z}\equiv x_+$. The WZW field
$h=h(z,\overline{z})$ is a spinless primary field whose conformal
dimension 
$$
\Delta_h\> =\> \overline{\Delta}_h\> =\> {c_h/2 \over
k + h_{g}^\vee}\>;
\nfr{AnDim}
depends on the chosen representation of $G$ through $c_h$, the
value of the quadratic Casimir of $G$, $t^a t^a= -c_h\> I$. The 
relation between the WZW field and the chiral currents is summarized
by the ``null-equation''
$$
\hbar \left(k+ h_{g}^\vee\right) \partial h\> =
\> \bigl(J^a\> t^a h\bigr)\>,
\nfr{Null}
where $(AB)(z)$ is the normal ordered product of two operators $A(z)$
and $B(z)$~[\Ref{BAIS1}]. 

In the last equations, the $t^a$'s provide an
antihermitian basis for the (compact) Lie algebra $g$ normalized such
that $\langle t^a, t^b\rangle =-\delta^{a,b}$. We will
distinguish the generators of the Cartan subalgebra
$g_{0}^0$ as
$t^A, t^B,\ldots$ with $A, B, \ldots=1,\ldots, r_g$, while the other
generators will be denoted as $t^i, t^j, t^k, \ldots$. The standard
realization of this basis by means of a Chevalley basis for the
complexification of $g$ is presented in the appendix.

Then, the operator algebra of the 
$G/U(1)^{\times r_g}$ coset CFT is generated by those operators whose OPE
with the currents $J^A(z)$ and $\overline{J^A}(\overline{z})$ is
regular~[\Ref{BAIS2}]. This condition is the
quantum version of gauge invariance with respect to~\Gtrans, as will 
be shown in the next section.

In eq.~\QAct, $\Phi =\langle  \Lambda_+,  h^{\dagger}
\Lambda_-  h\rangle$ will be understood as a matrix element of
the WZW field $h$ taken in the adjoint representation, $\Phi
=\langle \Lambda_-, h^{\rm ad}\cdot \Lambda_+\rangle$. Recall that,
with our normalization of
$\langle\>, \>\rangle$, the quadratic Casimir in the adjoint
representation is $c_v = \> f^{abc} f^{abc}\> =\> 2h_{g}^\vee$.
Therefore, the perturbation is given by a spinless primary field with
conformal dimension
$$
\Delta_\Phi\> =\> \overline{\Delta}_\Phi\> =\> {h_{g}^\vee \over
k + h_{g}^\vee}\> <\> 1\>,
\nfr{CDim}
which shows that the perturbation is relevant for any value of $k$.
Correspondingly, $m$ is a dimensionful coupling constant with positive
scale dimension
$$
\bigl[ m\bigr]\> =\> {k \over k + h_{g}^\vee}\> >\> 0\>.
\nfr{DimM}
All this is equivalent to the statement that the resulting field 
theory is ``super-renormalizable'', which means that only a finite
number of couterterms are required to renormalize the theory.
Moreover, we will assume that $2\> \Delta_\Phi\le 1$ or,
equivalently, $k\ge h_{g}^\vee$, and, consequently, that no
counterterm is actually needed~[\Ref{ZAM},\Ref{CARDY}].

It is worth considering the semi-classical and/or weak coupling
$k\rightarrow \infty$ limit of the HSG theories. Then, $C_{\rm CFT}
(G,k) \rightarrow {\rm dim}(G) - {\rm rank}(G)$ and $[ m] \rightarrow
1$, which shows that, in this limit, the theory consists of ${\rm
dim}(G) - {\rm rank}(G)$ free massive bosonic particles whose mass is
proportional to
$\hbar m$. They are just the ``fundamental particles'' of the
classical theory, which are associated to the roots of the Lie algebra
$g$~[\Ref{MASS}].

\chapter{Classical integrability}

At the classical level, the integrability of the HSG theories is
manifested in the zero-curvature form of their equations-of-motion
$$
\left[\partial_-\> +\> m\Lambda_-\> -\> \partial_-h\> h^\dagger\> +\>
h\tau(A_-) h^\dagger\> ,\> \partial_+\> +\> m h \Lambda_+
h^\dagger\> +\> A_+\right]\> =\>0 \>,
\nfr{ZeroC}
along with the constraints that arise from the equations of motion of
the gauge field:
$$
\eqalign{
&\bigl\langle t^A\> ,\> h^\dagger \partial_+h \> +\> h^\dagger A_+ 
h\> - \> \tau(A_+) \bigr\rangle\> =\>0 \>, \cr
&\bigl\langle t^A\> ,\> -\partial_- h h^\dagger \> +\>
h \tau( A_-) h^\dagger\> -\>
A_- \bigr\rangle\> =\>0\>, \cr}
\nfr{Const}
for all $A=1,\ldots, r_g$. Eq.~\ZeroC\ implies that there exist an
infinite number of conserved densities that can be obtained by 
applying the well known Drinfel'd-Sokolov construction~[\Ref{DS}].
This requires to express eq.~\ZeroC\ as a zero-curvature equation
associated to the ``homogeneous gradation'' of the loop algebra (see
the last reference in~[\Ref{DS}])
$$
{\cal L}(g)\>=\>{\Bbb C}[\lambda,\lambda^{-1}]\otimes g\> =\>  
\bigoplus_{j\in{\Bbb Z}} {\cal L}(g)_j\>,
\quad {\rm where}
\quad {\cal L}(g)_j\> =\> \lambda^j \otimes g\>,
\nfr{LoopA}
which is equivalent to the introduction of a ``spectral parameter'' 
$\lambda$. Since the equations of motion remain unchanged if
$\Lambda_\pm
\mapsto \lambda^{\mp1}\otimes \Lambda_\pm$, eq.~\ZeroC\ can actually 
be understood as a zero-curvature equation associated to ${\cal
L}(g)$ and to the Lax operator $L=\partial_- +\Lambda +q$, where
$$
\Lambda \> \equiv \>m \>\lambda \otimes \Lambda_-  \in  {\cal L}(g)_1
\quad{\rm and}
\quad q= -\> \partial_-h\> h^\dagger\> +\>
h\tau(A_-) h^\dagger \in {\cal L}(g)_0\>.
\nfr{LaxO}

In this case, the Drinfel'd-Sokolov construction goes as follows. 
First, there is some function $y$ of the form
$$
y\> =\> \sum_{n>0} \>y^i(n)\> \lambda^{-n}\otimes t^i\in {\cal
L}(g)_{<0}
\efr
that ``abelianizes'' the Lax operator, {\it i.e.}, 
$$
\eqalign{
& {\rm e\/}^{ y}\>\left( \partial_-\> +\>\Lambda\> +\> 
q\right) \>{\rm e\/}^{-y} \> =\> \partial_- \> + \>\Lambda \>
+\> H \>, \cr
& {\rm e\/}^{ y}\> \bigl( \partial_+\> +\>m\> h\> (\lambda^{-1}
\otimes
\Lambda_+)\> h^\dagger\>+\> A_+\bigr) \>{\rm e\/}^{-y}
\> =\> \partial_+ \> + \> \overline{H}\>, \cr}
\nfr{Abel}
where $H$ and $\overline{H}$ take values in the centralizer of
$\Lambda$ in ${\cal L}(g)_{\leq0}$: 
$$
H\>= \> \sum_{s\geq1}\> I_{s}^{(0)A}\> \lambda^{-s+1}\otimes t^A
\quad{\rm and}\quad 
\overline{H}\>= \> \sum_{s\geq1}\> \overline{I}_{s}^{(0)A}\>
\lambda^{-s+1}\otimes t^A\>.
\efr
Moreover, $y$, $H$, and $\overline{H}$ are local functionals of $q$,
and $I_{s}^{(0)A}$ and $\overline{I}_{s}^{(0)A}$ have scale 
dimension~$s$ with respect to the scale transformations $x_\pm
\rightarrow x_\pm/\rho$. Then, the zero curvature equation~\ZeroC\
simplifies to
$$
\partial_+ I_{s}^{(0)A}\> =\>\partial_- \overline{I}_{s}^{(0)A}\>,
\nfr{ClasC}
which shows that, for each scale dimension $s\geq1$, there are ${\rm
rank}(G)$ classically conserved densities $(I_{s}^{(0)A}\>, \>
\overline{I}_{s}^{(0)A})$.

The transformation of the conserved densities with respect
to the gauge transformations~\Gtrans\ can be easily derived by 
realizing that the Lax operator transforms just by conjugation:
$L\mapsto {\rm e\>}^\alpha\> L \> {\rm e\>}^{-\alpha}$ or,
%%%
%%%LUIS: I have modified the footnote.
%%%
equivalently,\note{Notice that the gauge transformation of the Lax
operator is independent of $\tau$, {\it i.e.\/}, it does not depend on
the the precise form of the group of gauge transformations~\Gtrans\ of
the theory. Actually, this is the reason why the proof of integrability
does not make any reference to the automorphism~$\tau$.}  
$$
q\mapsto {\rm e\>}^\alpha\> q \> {\rm e\>}^{-\alpha}\> -\>
\partial_-\alpha\>.
\nfr{Gaugeq} 
Then,
${\rm e\>}^y \mapsto {\rm e\>}^\alpha {\rm e\>}^y {\rm e\>}^{-
\alpha}$,
$H\mapsto H -\partial_-\alpha$, and $\overline{H}\mapsto   
\overline{ H}- \partial_+\alpha$, which means that all the conserved
densities with scale dimension $s\geq2$ are gauge invariant, while
the densities with $s=1$ transform as 
$$
I_{1}^{(0)A} \mapsto I_{1}^{(0)A}-\partial_-\alpha^A \quad {\rm and} 
\quad \overline{I}_{1}^{(0)A} \mapsto \overline{I}_{1}^{(0)A} -
\partial_+ \alpha^A\>.
\nfr{ConI}  
Actually, using eqs.~\Const\ and~\Abel, one can easily show that 
$I_{1}^{(0)A} \>t^A =A_-$ and $\overline{I}_{1}^{(0)A}\> t^A= A_+$, 
in agreement with eqs.~\Gtrans\ and~\ConI. 

The global symmetries that give rise to the $s=1$ conserved densities 
and the associated gauge invariant conserved charges have been
discussed in~[\Ref{MASS}]. In the following, we will only be
interested in the gauge invariant densities ($s\geq2$), and we provide
explicit expressions for the classically conserved densities with scale
dimension $s=2$ and~$3$ in  the appendix.

\chapter{Quantum conserved densities}

In this section we examine the fate of the classical conserved
quantities of spin 2 and 3 in the quantum theory. The proof that, 
after an appropriate renormalization, 
the $r_g$ currents of spin 2 and 3 remain conserved 
is quite involved and uses conformal perturbation theory. 

Since the quantum HSG theories can be described as perturbed conformal
field theories, the existence of quantum conserved charges can be
investigated by using the method of Zamolodchikov~[\Ref{ZAM}] (see
also~[\Ref{CARDY}]). In the presence of
the perturbation~\QAct\ any local chiral field
${\cal I}(z)$, which in the unperturbed CFT satisfies
$\overline{\partial} {\cal I}(z)=0$, adquires a $\overline{z}$ 
dependence:
$$
\overline{\partial}{\cal I}(z,\overline{z})\>=\> -\> k\>m^2 \>
\oint_z {dw\over 2\pi i}\> \Phi(w,\overline{z})\> {\cal I}(z)\>.
\nfr{MasterZam}
This is the contribution at the lowest order in perturbation theory;
however, if $k\geq h_{g}^\vee$, or equivalently $2\Delta_\Phi\leq1$,
this expression is expected to be exact~[\Ref{CARDY}] (see the
comments below eq.~\DimM).

Consequently, in the perturbed theory, ${\cal I}(z,\overline{z})$ 
will  be a conserved density if the right-hand-side of~\MasterZam\ 
is a total $\partial$ derivative, {\it i.e.}, if~\MasterZam\ can be
written as $\overline{\partial}{\cal I}(z,\overline{z}) =
\partial\overline{{\cal I}}(z,\overline{z})$ where $\overline{{\cal
I}}(z,\overline{z})$ is some field of the original CFT. For this to
be true, it is enough that the residue of the simple pole in the OPE
between ${\cal I}(z)$ and $\Phi(w,\overline{z})$ is a total
derivative,
$$
\Phi(w,\overline{z})\> {\cal I}(z) \>=\> \sum_{n>1}  {\{\Phi{\cal
I}\}_n \over (w-z)^n} \>+\> {1\over w-z}\> \partial \overline{{\cal
I}}(z,\overline{z}) \>+\> \cdots\>.
\nfr{Residue}
Recall that the residues of the simple pole in the OPE's $\Phi(w,
\overline{z} )\> {\cal I}(z)$ and ${\cal I}(z)\> \Phi(w, \overline{w})$
differ only in a total $\partial$ derivative and, in fact, we will
always consider the latter because its expression is usually simpler. 

Since the classical theory exhibits gauge invariant conserved densities
for any scale dimension $s\geq2$, we will study local composite
operators ${\cal I}_s$ of the chiral currents $J^a(z)$ and their
derivatives of conformal spin $s\geq2$. However, since the unperturbed
CFT is a coset CFT, these operators have to be constrained by the
condition of having a regular OPE with the currents $J^A(z)$, for $A=1,
\ldots, r_g$. Actually, this condition is the quantum version of gauge 
invariance in the classical theory, as shown by the OPE's
$$
\eqalign{
&J^A(z)\> J^B(w)  \>= \> - \hbar^2 k{\delta^{AB} \over (z-w)^2} \> +\>
\cdots \>, \cr
&J^A(z)\> J^i(w)  \>= \> \hbar\> {f^{Ai\>\overline{\imath} }\>
J^{\overline{\imath}}(w)\over z-w} \>+ \> \cdots\>,\cr}
\efr
compared with the infinitesimal gauge transformation~\Gaugeq\ of the
components of
$q$,
$$
\eqalignno{
& \delta q^A(z,\overline{z}) \>= \> -\partial \alpha^A(z,\overline{z})
\>=\> -\>
\oint_z {dw\over 2\pi i}\>\alpha^B(w,\overline{z})\> {\delta^{AB}\over
(w-z)^2} \>\cr 
\noalign{\vskip0.2cm}
& \delta q^i(z,\overline{z}) \>= \> -\> \alpha^A(z,\overline{z})\>
f^{Ai\>\overline{\imath}}\>
 q^{\overline{\imath}}(z,\overline{z}) 
\>=\> -\>
\oint_z {dw\over 2\pi i}\>\alpha^A(w,\overline{z})\> { f^{Ai\>
\overline{\imath}}\>
q^{\overline{\imath}}(w, \overline{z}) \over w-z}\>. &\numali \cr}
$$

At this point it is convenient to recall that classical expressions
can be recovered from quantum expressions through (see eqs.\Null\
and~\LaxO) 
$$
J^a t^a \>= \> - (\hbar k)\> q\>, \quad (\hbar k)\rightarrow
{1\over \beta^2}\>, \quad {\rm and} \quad k\rightarrow\infty\>.
\nfr{CLimit}

\section{Quantum spin-2 conserved densities}

We consider a generic (normal ordered) spin-2 operator
$$
{\cal I}_2(z) \> = \> {\cal D}_{ab} \> \bigl(J^a J^b\bigr)(z)\>,
\nfr{QStwo}
with a c-number tensor 
${\cal D}_{ab}={\cal D}_{ba}$. The condition that the OPE
$J^A(z)\> {\cal I}_2(w)$ is regular implies that the only non vanishing
components of ${\cal D}_{ab}$ are ${\cal D}_{\alpha \alpha} = {\cal
D}_{\overline{\alpha}\>  \overline{\alpha}}$ for any positive root
$\alb$,\note{${\cal D}_{\alpha \alpha}$, ${\cal D}_{\overline{\alpha}\> 
\overline{\alpha}}$, and ${\cal D}_{AB}$ are components of
$D_{ab}$ with respect to the basis~(A.2), which will be used throughout
this Section.} and
$$
{\cal D}_{AB}\> =\> -\>{1\over k} \sum_{\alb>0} {\cal D}_{\alpha 
\alpha}\>
\alpha^A\> \alpha^B\>,
\nfr{GQStwo}
which shows that ${\cal D}_{AB}$ vanishes
in the $k\rightarrow \infty$ limit and therefore is a quantum
correction.

To investigate the condition that ${\cal I}_2$ is a classical density,
we use the OPE
$$
J^a(z)\> h(w,{\overline w})\> = \> -\> \hbar \> {t^a\> h(w,{\overline w})
\over z-w}\> +\> \cdots\>,
\nfr{OPEJh}
which is satisfied in an arbitrary representation of $G$. In
particular, by considering the adjoint representation, eqs.~\Null\
and~\OPEJh\ lead to
$$
\eqalignno{
& \hbar \left(k+ h_{g}^\vee\right) \partial\> \langle v, P \rangle \>
= \> \bigl(J^a\> \langle [v,t^a]\>, \>P\rangle \bigr)\>,
&\nameali{OPEadja}\cr
\noalign{\vskip0.2cm}
&J^a(z)\> \langle v\>, \>P(w,{\overline w})\rangle\> = \> -\> \hbar \>
{\langle  [v,t^a]\>, \>P(w,{\overline w})\rangle
\over z-w}\> +\> \cdots\>, &\nameali{OPEadjb}\cr
}
$$
where $v$ is an arbitrary element in $g$, $P = \bigl( h \Lambda_+
h^\dagger \bigr)\equiv h^{\rm ad}\cdot \Lambda_+$, and the perturbing
operator is $\Phi = \langle
\Lambda_-,P\rangle$. Using this, it is easy to show that the
residue of the simple pole in the OPE ${\cal I}_2(z)
\Phi(w,\overline{w})$ is 
$$
{\rm Res}\Bigl({\cal I}_2(z) \Phi(w,\overline{w})\Bigr)\> =
\> -\ 2\hbar\> \bigl(J^a\> \langle {\cal D}_{ab}\>[\Lambda_-, t^b]\> 
,\> P\rangle\bigr)(w,\overline{w})\>.  
\nfr{ResStwo}
Therefore, using~\OPEadja, ${\cal I}_2$ gives rise to a quantum 
%%%
%%%LUIS: I have incorporated in the footnote the comment you added
%%%after the next equation.
conserved density if ${\cal D}_{ab}$ is chosen such that~\note{The
normalization is chosen to simplify the comparison with the classical
results. Explicit factors of $\hbar k$ arise because of the relation
$J^a =  -(\hbar k)\> q^a$ (see eq.~\CLimit).}
$$
(\hbar k)^2\> {\cal D}_{ab}\>[\Lambda_-, t^b]\> =\>
{1\over2}\>[\mub\cdot {\bfmath t}, t^a]\>,
\efr
for any $r_g$-component vector $\mub$, which leads to
$$
\eqalignno{
& (\hbar k)^2\>{\cal D}_{\alpha \alpha}(\mub) \>= \>(\hbar k)^2\>
{\cal D}_{\overline{\alpha}\>
\overline{\alpha}}(\mub)\> =\> {1\over2}\> {\mub\cdot\alb \over 
\lab\cdot\alb} \>= \> m\> D_{\alpha \alpha}^{(0)}(\mub) \>, \cr
\noalign{\vskip0.2cm}
& (\hbar k)^2\>{\cal D}_{AB}(\mub) \>= \> -\>{1\over 2k} \sum_{\alb>0}
{\mub\cdot\alb
\over \lab\cdot\alb}\> \alpha^A\> \alpha^B\>, &\nameali{StwoFin} \cr}
$$
where $D_{\alpha \alpha}^{(0)}= D_{\overline{\alpha}\>
\overline{\alpha}}^{(0)}$ gives the classical conserved density, 
$\Lambda_- = \lab\cdot {\bfmath t}$ (see the appendix), and the value
of ${\cal D}_{AB}$ follows from eq.~\GQStwo.

Therefore, we conclude that there exists ${\rm rank}(G)$ quantum
conserved densities of scale dimension $s=2$ labelled by the arbitrary
vector $\mub$:
$$
\eqalignno{
&(\hbar k)^2\> {\cal I}_2(\mub)\> =\> {1\over2}\> \sum_{\alb>0}
{\mub\cdot\alb
\over \lab\cdot\alb}\> \Bigl[
\bigl(J^\alpha J^\alpha\> +\> J^{\overline{\alpha}}
J^{\overline{\alpha}}\bigr)\> -\> {1\over k}\bigl(J^A J^B\bigr)\>
\alpha^A \alpha^B \Bigr]\>, \cr
\noalign{\vskip0.2cm}
&(\hbar k)^2\> \overline{{\cal I}}_2(\mub)\> =\> -\> m^2 \> (\hbar
k)^2\> \langle
\mub\cdot {\bfmath t}\> ,\> P\rangle\>. & \nameali{Goodtwo}\cr}
$$
The relation between quantum and classical conserved densities of 
scale dimension~2 is obtained by considering~\CLimit,  
$$
\eqalignno{
&{\cal I}_2(\mub) \> =\> m\>\mub\cdot\left({\bfmath
I}_{2}^{(0)}
\> +\>
{1\over k}\> {\bfmath I}_{2}^{(1)} \right)\>, \cr
&\overline{{\cal I}}_2(\mub) \> =\> m \> \mub\cdot
\overline{{\bfmath I}}_{2}^{(0)}\>. & \numali \cr}
$$

Finally, let us exhibit the relation between the quantum conserved
density ${\cal I}_2(\lab)$ and the energy-momentum tensor. First, 
recall the algebraic relation
$$
\sum_{\alpha>0} \alpha^A \alpha^B \> =\> {c_v\over 2}\>
\delta^{A,B}\> = \> h_{g}^\vee \> \delta^{A,B}\>.
\efr
This implies that eq~\Goodtwo\ becomes
$$
(\hbar k)^2\> {\cal I}_2(\lab) \> =\> -(k+ h_{g}^\vee) \Bigl\{
{-1\over 2(k+ h_{g}^\vee)} \> \bigl(J^a J^a\bigr)\> +\>
{1\over 2k}\> \bigl(J^A J^A\bigr) \Bigr\} \>\equiv \>
-(k+ h_{g}^\vee)\> T_{z,z}\>,
\nfr{Tzz}
in agreement with the Sugawara construction of the energy-momentum 
tensor of the coset CFT, and, consequently,
$$
(\hbar k)^2\> \overline{{\cal I}}_2(\lab) \> =\> -\> m^2 \> (\hbar
k)^2\> \langle\Lambda_- \> ,\> P\rangle \>\equiv \>
-(k+ h_{g}^\vee)\> T_{z,\overline{z}}\>.
\nfr{Tzzbar}      
  
\section{Quantum spin-3 conserved densities}

Let us consider a (normal ordered) spin-3 operator of the form
$$
\eqalignno{
{\cal  I}_3 &(z)\>  =\>
{\cal P}_{ijk}\> \bigl(J^i\> \bigl(J^j J^k\bigr) \bigr)(z)\> +\> 
{\cal P}_{Aij}\> \bigl(J^A\> \bigl(J^i J^j\bigr) \bigr)(z)\> +\>
{\cal Q}_{ij}\> \bigl(J^i \partial J^j \bigr)(z) \cr
\noalign{\vskip0.2cm}
& +\> {\cal P}_{ABC}\> \bigl(J^A\> \bigl(J^B J^C\bigr) \bigr)(z)\> 
+\> {\cal Q}_{AB}\> \bigl(J^A \partial J^B \bigr)(z) \>
+\> {\cal R}_A\> \partial^2 J^A(z)\>, & \nameali{ThreeGen} \cr}
$$
where ${\cal P}_{ijk}$ and ${\cal P}_{ABC}$ are totally symmetric,
${\cal P}_{Aij}= {\cal P}_{Aji}$, and ${\cal Q}_{ij}$ and ${\cal
Q}_{AB}$ are antisymmetric c-number tensors. 

The requirement of gauge invariance, {\it i.e.\/}
that the  OPE's $J^A(z)\> {\cal I}_3(w)$ are regular
implies the following constraints. Firstly,
$$
\eqalignno{
& {\cal P}_{ABC}\> =\> -\> {1\over 6k}\> \left( f^{Aik}\> f^{Bjk}\>
{\cal P}_{Cij}\> + \> f^{Aik}\> f^{Cjk}\>
{\cal P}_{Bij} \right)\>, \cr
& {\cal Q}_{AB}\> = \>- \> {\hbar\over 2k}\> f^{Ail}\> f^{jlm}\>
f^{Bkm}\> {\cal P}_{ijk}\>, \quad {\rm and} \quad
{\cal R}_A\> =\> {\hbar\over6}\> f^{Aij}\> {\cal Q}_{ij} \>, &
\nameali{ConstOne}\cr}
$$
which, taking into account the relation between quantum and
classical variables, $J^a= -(\hbar k) q^a$, can be written as 
$$
\eqalignno{
& [-(\hbar k)^3{\cal P}_{ABC}]\> =\> -\> {1\over 6k}\> \Bigl( 
f^{Aik}\> f^{Bjk}\> [-(\hbar k)^3{\cal P}_{Cij}]\> + \> f^{Aik}\>
f^{Cjk}\> [-(\hbar k)^3{\cal P}_{Bij}] \Bigr)\>, \cr
& [(\hbar k)^2{\cal Q}_{AB}]\> = \>+ \> {1\over 2k^2}\> f^{Ail}\>
f^{jlm}\> f^{Bkm}\> [-(\hbar k)^3{\cal P}_{ijk}]\>,\cr 
& [-(\hbar k){\cal R}_A]\> =\> -\> {1\over6k}\> f^{Aij}\> [(\hbar
k)^2{\cal Q}_{ij}] \>, & \nameali{ConstOnePlus}\cr}
$$
which shows that ${\cal P}_{ABC}$, ${\cal Q}_{AB}$, and  ${\cal
R}_A$ arise as quantum corrections. Secondly, the only non-vanishing
components of
${\cal P}_{Aij}$ and
${\cal Q}_{ij}$ are ${\cal P}_{A\alpha \alpha} = {\cal P}_{A
\overline{\alpha}\> \overline{\alpha}}$ and $ {\cal Q}_{\alpha
\overline{\alpha}} = - {\cal Q}_{ \overline{\alpha} \alpha}$, for any
positive root $\alb$. Finally,
$$
\alpha^A\> {\cal Q}_{\alpha \overline{\alpha}} \> +\> (\hbar k) \>
{\cal P}_{A\alpha \alpha} \> =\> -\> 3\hbar\> f^{A i k}\> f^{j k
\alpha}\> {\cal P}_{ij\alpha}\>,
\nfr{ConstTwo}
as can be checked by means of an involved but straightforward
calculation.

To study the quantum conservation of~\ThreeGen, it is more
convenient to write the spin-3 density as
$$
{\cal I}_3(z)\> =\> 
\widehat{\cal P}_{abc}\> \bigl(J^a\> \bigl(J^b J^c\bigr) \bigr)(z)\>
+\>
\widehat{\cal Q}_{ab}\> \bigl(J^a \partial J^b \bigr)(z)
+\> \widehat{\cal R}_A\> \partial^2 J^A(z)\>,
\nfr{ThreeGen}
where $\widehat{\cal P}_{abc}$ is totally symmetric and 
$\widehat{\cal Q}_{ab}$ is antisymmetric, and we have the following
identities (recall that we are using the conventions of~[\Ref{BAIS1}]
for normal ordered products)
$$
\eqalign{
&{\cal P}_{ijk} \>= \> \widehat{\cal P}_{ijk}\>, \qquad
{\cal P}_{Aij}\> =\> 3\> \widehat{\cal P}_{Aij} \>, \cr
&{\cal Q}_{ij} \>= \> \widehat{\cal Q}_{ij}\> -\>
{3\hbar\over 2}\> \left(f^{Aik}\> \widehat{\cal P}_{Ajk}  \>- \>
f^{Ajk}\> \widehat{\cal P}_{Aik}\right)\>. \cr}
\nfr{Connect}
In this way, it is easy to show that
$$
{\rm Res}\Bigl({\cal I}_3(z) \Phi(w,\overline{w})\Bigr)\> =
\> \hbar\Bigl\{ \bigl(J^a\> \bigl( J^b \> \langle X_{ab}\>, \>
P\rangle \bigr)\bigr) \> +\> \bigl( \partial J^a\> \langle
\Omega_a \>, \> P\rangle \bigr) \Bigr\}(w,\overline{w})\>,
\nfr{OPEthree}
where the algebraic coefficients are
$$
\eqalign{
& X_{ab}\> =\> -\> 3\> \widehat{\cal P}_{abc}\> [\Lambda_-\>, \> t^c]
\>, \cr 
& \Omega_a\> =\> 2\> \widehat{\cal Q}_{ab}\> [\Lambda_-\>, \>
t^b]\> +\> 3\hbar \> \widehat{\cal P}_{abc}\> [ [\Lambda_-\>, \>
t^b]\>, \> t^c]\>. \cr}
\nfr{Tensors}

Therefore, taking into account~\OPEadja, the condition to have a spin-3
quantum conserved density is that
$$
\eqalign{
& X_{ab}\> =\> [F_a\>, \> t^b]\> +\> [F_b\>, \> t^a]\>, \cr
& \Omega_a \>= \> 2\hbar\> (k + h_g^\vee)\> F_a \>+ \> \hbar \>
f^{abc}\> [F_b\>, \> t^c]\>, \cr} 
\nfr{ConsCond}
for some elements $F_a$ in $g$, which ensures that
$$
{\rm Res}\Bigl({\cal I}_3(z) \Phi(w,\overline{w})\Bigr)\> =
\> 2\hbar^2\> (k + h_g^\vee)\> \partial \bigl(J^a\> \langle F_a \>,
\> P\rangle \bigr) (w,\overline{w}) \>.
\nfr{TotDerthree}
Eqs.~\ConsCond\ can be solved for $\widehat{\cal P}_{abc}$ and
$\widehat{\cal Q}_{ab}$ as functions of $F_a$:
$$
\eqalignno{
& 3 \> \widehat{\cal P}_{abc}\> [\Lambda_-\>, \> t^c]\> =\>
-\> [F_a\>, \> t^b]\>- \> [F_b\>, \> t^a]\>, &\nameali{Solutiona}\cr
& 2 \> \widehat{\cal Q}_{ab}\> [\Lambda_-\>, \> t^b]\> =\>
2(\hbar k)\> F_a\> +\> \hbar\> \Bigl\{ [F_b\>, \> [t^a\>, \> t^b]]\>
+\> [t^b\>, \> [t^a\>, \> F_b]] \Bigr\}\>. &\nameali{Solutionb}\cr}
$$

The condition to have a classical conserved density of scale
dimension~3 is recovered by removing the last term in eq.~\Solutionb,
which is an explicit quantum correction. Then, the resulting classical
equations have ${\rm rank}(G)$ solutions
$$
(\hbar k)^3 \> F_{a}^{(0)}\> =\> m^2\> Q_{ab}^{(0)}(\mub)\> 
[\Lambda_-, t^b]\>,
\efr
which give rise to the ${\rm rank}(G)$ spin-3 classically conserved
densities presented in the appendix. Notice that, for these classical
solutions, $F_A \equiv F_{A}^{(0)} =0$. 

The quantum corrections to these classical solutions are induced by 
the last term in eq.~\Solutionb. In fact, this equation with $a=A$
becomes a constraint for $F_a$
$$
2(\hbar k)\> F_A\> +\> \hbar\> \Bigl\{ [F_b\>, \> [t^A\>, \> t^b]]\>
+\> [t^b\>, \> [t^A\>, \> F_b]] \Bigr\}\>= 0\>, 
\efr
which admits the exact solution
$$
F_i \> = \> F_{i}^{(0)} \>, \quad
F_{A}\> =\> -\> {1\over 2k}\> \Bigl\{ [F_{i}^{(0)}\>, \> [t^A\>, \>
t^i]]\> +\> [t^i\>, \> [t^A\>, \> F_{i}^{(0)}]] \Bigr\}\>,
\efr
which, therefore, can be understood as the ``renormalization'' of the
classical solution $F_a = F_{a}^{(0)}$.

Finally, eqs.~\Solutiona, \Solutionb, and~\Connect\ provide the
precise value of the tensors that specify the conserved spin-3 density
${\cal I}_3$, namely
$$
\eqalignno{
-\> &(\hbar k)^3\> {\cal P}_{ijk}\> =\> m^2\> P_{ijk}^{(0)} (\mub) \>,
& \nameali{ThreeIndex} \cr
\noalign{\vskip0.2cm}
-\> & (\hbar k)^3\> {\cal P}_{Aij}\> =\> m^2\> \left( P_{Aij}^{(0)}
(\mub)\> +\> {1\over k}\> P_{Aij}^{(1)}(\mub) \right)\>,
& \nameali{TwoIndex} \cr
\noalign{\vskip0.2cm}
& (\hbar k)^2\> {\cal Q}_{ij}\> =\> m^2\> \left( Q_{ij}^{(0)}
(\mub)\> +\> {1\over k}\> Q_{ij}^{(1)}(\mub) \right)\>,
& \nameali{TwoDer} \cr}
$$
where the classical contributions can be found in the appendix.
The quantum corrections are given by
$$
\eqalignno{
& P_{A\alpha \alpha}^{(1)}(\mub) \> = \> 2\> \sum_{\beb>0} \>
{(\lab\cdot \beb)\> (\alb\cdot \beb)\over (\lab\cdot\alb)} \>
\beta^A\> Q_{\beta \> \overline{\beta}}^{(0)} (\mub)\>, &
\nameali{TwoQuan}\cr 
& Q_{\alpha\> \overline{\alpha}}^{(1)}(\mub) \> = \> \alpha^A\>
P_{A\alpha \alpha}^{(0)} \> +\>  \sum_{\beb>0} \> {(\lab\cdot
\beb)\over (\lab\cdot\alb)}\> D(\beb, \alb)\> Q_{\beta \>
\overline{\beta}}^{(0)} (\mub)\>, &
\nameali{DerQuan}\cr}
$$
and we have introduced the notation
$$
D(\beb, \alb) \> = \> \sum_{i} \bigl( f^{\alpha\beta i}\>
f^{\alpha\beta i}\> +\>  f^{\alpha\overline{\beta} i}\>
f^{\alpha\overline{\beta} i}\bigr)\>,
\efr
such that
$$
[t^\beta,[t^\alpha,t^\beta]]\> +\> [t^{\overline{\beta}},
[t^\alpha,t^{\overline{\beta}}]]\>=\> D(\beb, \alb) \> t^\alpha\>.
\efr  

By means of a tedious but straightforward calculation, it can be shown
that the coefficients in eqs.~\ThreeIndex-\TwoDer\ satisfy
the constraints~\ConstTwo, as required to ensure that ${\cal I}_3$ is
a generator of the coset operator algebra. Correspondingly,
eq.~\TotDerthree\ leads to
$$
\eqalignno{
& \overline{\cal I}_3(\mub)\> =\> -\>(\hbar k)\> m^2\> \bigl(
J^a\langle \widetilde{F}_a\>, \> P\rangle\bigr)\>, \cr
& \widetilde{F}_a\> =\> -\> 2\hbar\> ( k+ h_{g}^\vee)\> F_a \>+ \>
\widehat{\cal Q}_{ab}\> [\Lambda_-\>,\> t^b]\> + \> 3\hbar\> 
\widehat{\cal P}_{abc}\> [[\Lambda_- \>,\>t^b]\>,\> t^c]\>. &\numali
\cr}
$$
Therefore, for any simple compact Lie group $G$, we conclude that the
classically conserved densities give rise to precisely ${\rm rank}(G)$
quantum conserved densities 
$$
\eqalignno{
&{\cal I}_3(\mub) \> =\> m^2
\>\mub\cdot\left({\bfmath I}_{3}^{(0)} \> +\>
{1\over k}\> {\bfmath I}_{3}^{(1)} \>+ \>
{1\over k^2}\> {\bfmath I}_{3}^{(2)} \right)\>, \cr
&\overline{\cal I}_3(\mub) \> =\> m^2
\>\mub\cdot\left(\overline{\bfmath I}_{3}^{(0)} \>
+\> {1\over k}\> \overline{\bfmath I}_{3}^{(1)} \>+ \>
{1\over k^2}\> \overline{\bfmath I}_{3}^{(2)} \right)\>, & \numali
\cr}
$$
of scale dimension~3 (see eqs.~\ConstOnePlus, and~\ThreeIndex-\TwoDer).

\chapter{Conclusions and Discussion}

%%%%
%%%%LUIS: I have added some reference to \tau here.
%%%%
We have shown that the Homogeneous sine-Gordon theories
of~[\Ref{MASS}] exhibit $r_g$ quantum conserved densities of scale
dimension~2 and~3, which implies that these theories have a factorized
$S$-matrix~[\Ref{THREE}] and, hence, that they are quantum integrable.
The quantum conserved currents are related to the classical ones by a
non-trivial renormalization.
These theories are perturbations of the level-$k$ theories of
$G$-parafermions, which are well-known conformal field
theories~[\Ref{GPARAF}], where $G$ is a compact simple Lie group.
They are characterized by an anomaly-free embedding of the
maximal abelian subgroup~$U(1)^{\times r_g}$ into $G_L\otimes G_R$,
which is specified by an automorfism $\tau$ of the Cartan subalgebra
that preserves the Killing form $\langle\>, \>\rangle$. Consequently,
$\tau$~can be (almost) any element of the orthogonal group~$O(r_g)$,
and hence it provides a set of continuous (marginal) coupling
constants. The perturbation is given by a relevant primary field
specified by two constant elements $\Lambda_\pm$ in the Lie algebra of
$G$, which play the role of continuous vector coupling
constants of the theory. In contrast, the other coupling constant in
the theory $\beta^2$ has to be quantized, giving rise to the integer
level~$k$,  if the quantum theory is to be well
defined~[\Ref{WITTEN}]. Our results show that the
resulting theories are quantum integrable for any compact simple Lie
group $G$, and for any value of the level $k$ (at least if $k\geq$ the
dual Coxeter number of $G$). More remarkably the theories are
integrable for any value of the vector coupling
constants $\Lambda_\pm$ and for any choice of the automorphism~$\tau$, a
fact which appears to be in contradiction to the commonly held belief
that integrability requires careful fine-tuning of coupling constants.

An important property of these theories is that they have soliton
solutions, like the sine-Gordon theory. As explained in~[\Ref{MASS}],
the solitons and fundamental particles of the HSG theories are
associated to the positive roots of
$g$, the Lie algebra of $G$. Recall that the simplest theory of this
class is the complex sine-Gordon theory
(CSG)~[\Ref{PARK1},\Ref{BAK},\Ref{TN}], which has periodic
time-dependent soliton solutions (see~[\Ref{TN}] and references
therein). For the more general theory, the solitons are constructed by
embeddings of the CSG soliton associated to a particular $su(2)$ subalgebra 
generated by $E_{\pm \alb}$.

The quantum integrability of these theories implies that they should
admit a factorizable $S$-matrix and the next stage of analysis
consists in establishing its form. We expect that it should be
possible to infer the form of the exact $S$-matrix through the
semi-classical quantization of the solitons.
Actually, these solitons carry abelian charges and the expectation is
that they give rise to a tower of massive
states where the lightest ones correspond to the fundamental
particles appearing in the Lagrangian, 
as in the complex sine-Gordon theory. More particularly,
the semi-classical limit of the S-matrix can be compared with the
time-delays that occur in the classical
scattering~[\Ref{JW}]. The classical and semi-classical analysis of
the soliton spectrum, together with the expressions for the
time-delays, will be presented in a subsequent publication.

Finally, let us recall that all these theories can be viewed as
generalizations of the perturbation of the simplest ${\Bbb Z}_k$
parafermions by their first thermal operator~[\Ref{BAK}], which is
only one of their three known series of massive integrable
perturbations~[\Ref{PINT}]. It would be interesting, therefore,
to investigate the existence of other massive integrable
perturbations of the theory of $G$-parafermions for a given compact
Lie group~$G$. In this respect, let us recall that, according
to~[\Ref{MASS}], the non-abelian affine Toda equations give rise to a
second family of classical massive solitonic integrable theories
associated to (compact) Symmetric-Spaces. The theory associated
to a symmetric space $G/G_0$ describes a perturbation of the 
CFT's corresponding to a coset of the form $G_0 /U(1)^{\times
p}$, where $p$ is always $< {\rm rank}(G)$ and $\geq {\rm rank}(G) -
{\rm rank}(G/G_0)$. Thus, we expect that the analysis of this second
class of models along the lines of this paper could lead to different
solitonic integrable perturbations of more general coset CFT's and, as
particular examples, of parafermionic ($p={\rm rank}(G_0)$) as well as
WZW ($p=0$) theories.

\bjump
\centerline{\bf Acknowledgements}

\jump
\noindent
C.R.F.P., M.V.G. and J.L.M. thank Joaqu\'\i n S\'anchez Guill\'en and
Alfonso~V. Ramallo for useful discussions. They are  supported
partially by CICYT (AEN96-1673) and DGICYT (PB93-0344). T.J.H. is
supported by a PPARC Advanced Fellowship and would like to thank
IBERDROLA for assistance at the University of Santiago de
Compostela, under their Science and Technology Programme. We are also
grateful for partial support from the EC Comission via a TMR Grant,
contract number FMRX-CT96-0012.

\bjump 

\appendix{Appendix: The classically conserved
densities}

Below, we give the explicit expressions for the the classical conserved
densities with scale dimension $s=2$ and $3$, which can be obtained by
solving eqs.~\Abel. 

First, let us consider an explicit realization of the basis $\{t^a\}$ 
in terms of a Chevalley basis of the complexification of the compact
Lie algebra $g$. It consists of a Cartan subalgebra with generators,
$H^A$, $A=1, \ldots, r_g$, and step operators $E_\alpha$
normalized such that
$$
[H^A, E_\alpha]\> =\> \alpha^A\> E_\alpha\>, \quad{\rm and} \quad
[ E_\alpha, E_{-\alpha}]\> =\> {2\over \alb^2}\>  \alb\cdot  {\bfmath
H}\> \equiv \> {2\over \alb^2}\> \alpha^A H^A \>.
\efr
Then, the antihermitian basis for the compact Lie algebra $g$ is
$$
\eqalign{
&t^A\> =\> iH^A\>, \quad A=1, \ldots, r_g\>, \cr
& t^\alpha\> =\> \sqrt{\alb^2\over 4}\> \bigl(E_\alpha \> -\>
E_{-\alpha}\bigr)\>, \quad{\rm and}\quad
t^{\overline{\alpha}\>}\> =\> i\> \sqrt{\alb^2\over 4} \bigl(E_\alpha 
\>  +\> E_{-\alpha}\bigr)\>, \cr}
\nfr{Basis}
for any positive root $\alb$ of $g$. This way, in particular,  $f^{
\overline{\alpha}\>\overline{\beta}\>\overline{\gamma}} =  f^{
\alpha\beta\overline{\gamma}}=0$ for any three positive roots
$\alb, \beb, \gab$, and 
$[t^\alpha, t^{\overline{\alpha}\>}] = \alb\cdot {\bfmath t}\equiv
\alpha^A t^A$, which means that $ f^{A \alpha
\overline{\alpha}} = \alpha^A$. With respect to this basis, $\Lambda_-=
\lab\cdot {\bfmath t}$ (recall that $\Lambda = m\> z\otimes 
\Lambda_-$), and we will also use the following notations. First,
$t^{\overline{\imath}}$ indicates $t^{\overline \alpha}$ when
$i=\alpha$, or $t^{\alpha}$ if $i= {\overline \alpha}$, and, second, 
for any
$r_g$-component vector $\mub$, $f^{\mu ij}\equiv \mu^A f^{Aij}$.

Introducing $q=q^a t^a$, the expression for the $s=2$ classical
conserved densities is
$$
\eqalign{
& \mu^A\> I^{(0)A}_2\> \equiv \mub\cdot {\bfmath I}^{(0)}_2 \> =\>
D^{(0)}_{i,j}(\mub)\> q^i q^j\>,\cr
& \mu^A\> \overline{I}^{(0)A}_2\> \equiv \mub\cdot \overline{{\bfmath
I}}_{2}^{(0)}\> = \> -\> m\> \langle \mub\cdot {\bfmath t}\>, \>
h\Lambda_+ h^\dagger\rangle\>, \cr}
\nfr{CStwo}
where the only non-vanishing components of $D^{(0)}_{i,j}(\mub)$ are
$$
D^{(0)}_{\alpha \alpha}(\mub)\> =\> D^{(0)}_{\overline{\alpha}\> 
\overline{\alpha}} (\mub)\> =\> {1\over2m} {\mub\cdot \alb\over
\lab\cdot \alb}\>.
\nfr{CStwoT}
Here, $\mub$ is an arbitrary vector which manifests the existence of
${\rm rank}(G)$ conserved densities for each integer scale 
dimension~$s$.

As expected, one of the $s=2$ densities is related to the classical
energy-momentum tensor, which is recovered for
$\mub =m\lab$~[\Ref{MASS}]
$$
\eqalignno{
m\>\lab\cdot {\bfmath I}_{2}^{(0)} \>& =\>  {1\over2} \sum_{\alb>0}
\left( q^\alpha q^\alpha\> +\>  q^{\overline \alpha} q^{\overline 
\alpha} \right)\> \cr 
& = \> - {1\over2} \langle q-q^A t^A\>, \> q-q^A
t^A\rangle\>\equiv \> 4\pi\beta^2\> T_{--}\>, \cr
\noalign{\vskip0.2cm}
m\>\lab\cdot \overline{{\bfmath I}}_{2}^{(0)} \>& =\> -\> m^2\> \langle
\Lambda_-\>, \> h\Lambda_+ h^\dagger\rangle\>\equiv \> 4\pi\beta^2\>
T_{+-}\>. &\nameali{CEMT} \cr}
$$ 

For scaling dimension $s=3$, the ${\rm rank}(G)$ classically conserved
densities are of the form 
$$
\mub\cdot {\bfmath I}_{3}^{(0)} \> =\> P_{ijk}^{(0)}(\mub)\> q^i q^j
q^k\> +
\> P_{Aij}^{(0)}(\mub)\> q^A q^i q^j \> +\> Q_{ij}^{(0)}(\mub) q^i
\partial_- q^j\>,
\nfr{CSthree}
where the only non-vanishing coefficients are
$$
\eqalignno{
& P_{ijk}^{(0)}(\mub)\> =\> {f^{\mu i\overline{\imath}}\> f^{\lambda
j\overline{\jmath}} -  f^{\mu j\overline{\jmath}}\> f^{\lambda i
\overline{\imath}}
\over 6m^2\>  f^{\lambda i\overline{\imath}} f^{\lambda
j\overline{\jmath}}  f^{\lambda k\overline{k}}}\> f^{i j
\overline{k}}\>,& 
\nameali{CSthreeTa} \cr
\noalign{\vskip0.2cm}
& P_{A\alpha\alpha}^{(0)}(\mub)\>=\> P_{A\overline{\alpha}\>
\overline{\alpha}}^{(0)}(\mub)\>=\>  -{1\over2m^2}\> {
\mub\cdot\alb\over (\lab\cdot\alb)^2}\> \alpha^A\>, &
\nameali{CSthreeTb} \cr  
\noalign{\vskip0.2cm}
& Q_{\alpha\overline{\alpha}}^{(0)}(\mub)\>=\> - Q_{\overline{\alpha}
\alpha}^{(0)}(\mub) \> = \> -{1\over2m^2}\> { \mub\cdot \alb \over
(\lab\cdot\alb)^2}\>. &
\nameali{CSthreeTc} \cr}
$$
Notice that $P_{A\alpha\alpha}^{(0)}(\mub)= \alpha^A
Q_{\alpha\overline{\alpha}}^{(0)}(\mub)$, which, according
to~\ConstTwo, is a manifestation of the gauge invariance of
$\mub\cdot {\bfmath I}^{(0)}_3$. Correspondingly,
$$
\mub\cdot \overline{{\bfmath I}}_{3}^{(0)}\> =\> - 
\> {1\over 2}\> \sum_{\alb>0} {(\mub\cdot \alb)\over (\lab\cdot
\alb)}\> \langle q^\alpha t^\alpha\> +\> 
q^{\overline{\alpha}} t^{\overline{\alpha}}\> ,\> h\Lambda_+
h^\dagger\rangle\>.   
\efr

\bjump
\references

\beginref
\Rref{DELIUS}{G.W.~Delius, M.T.~Grisaru, and D.~Zanon, Nucl. Phys.
{\bf B 385} (1992) 307.}
\Rref{MARG}{M.~Henningson and C.R.~Nappi, Phys. Rev. {\bf D 48} (1993)
861; \newline
S.F.~Hassan and A.~Sen, Nucl. Phys. {\bf B 405} (1993)
143;\newline
E.B.~Kiritsis, Mod. Phys. Lett. {\bf A 6} (1991) 2871;
Nucl. Phys. {\bf B 405} (1993) 109.}
\Rref{GAUGE}{E.~Witten, Commun. Math. Phys. {\bf 144} (1992)
189; \newline
K.~Gawedzki and A.~Kupianen, Nucl. Phys. {\bf B 320} (1989)
625.}
\Rref{JW}{R. Jackiw and G. Woo, Phys. Rev. {\bf D12} (1975) 1643}
\Rref{PINT}{V.A. Fateev, Int. J.~Mod. Phys. {\bf A 6} (1991) 2109;
\newline
V.A. Fateev and A.B. Zamolodchikov, Phys. Lett. {\bf B 271}
(1991) 91;\newline
L.~Palla, Phys. Lett. {\bf B 253} (1991) 342.}
\Rref{TN}{N.~Dorey and T.J.~Hollowood, Nucl. Phys. {\bf B 440} (1995)
215.}
\Rref{BAK}{I.~Bakas, Int. J.~Mod. Phys. {\bf A 9} (1994) 3443.}
\Rref{PARK1}{Q-H.~Park, Phys. Lett. {\bf B 328} (1994) 329.}
\Rref{HOLL}{T.J.~Hollowood, Nucl. Phys. {\bf B 384} (1992) 523.}
\Rref{NAT}{A.N.~Leznov and M.V.~Saveliev, Commun. Math. Phys. {\bf
89} (1983) 59; {\sl Group theoretical methods for integration of
non-linear dynamical systems\/}, Prog. Phys.~15 (Birkhauser, Basel,
1992); \newline
J.~Underwood, {\sl Aspects of Non-Abelian Toda Theories\/},
Imperial/TP/92-93/30, hep-th/9304156; \newline
L.A.~Ferreira, J.L.~Miramontes, and J.~S\'anchez
Guill\'en, Nucl. Phys. {\bf B 449} (1995) 631.}
\Rref{THREE}{S.~Parke, Nucl. Phys. {\bf B 174} (1980) 166.}
\Rref{GPARAF}{A.B.~Zamolodchikov and V.A.~Fateev, Sov. Phys. JETP
{\bf 62} (1985) 215; \newline
D.~Gepner, Nucl. Phys. {\bf B 290} (1987) 10; \newline
D.~Gepner and Z.~Qiu, Nucl. Phys. {\bf B 285} (1987) 423.}
\Rref{DS}{V.G.~Drinfel'd and V.V.~Sokolov, J.~Sov. Math. {\bf 30}
(1985) 1975; Soviet. Math. Dokl. {\bf 23} (1981) 457;\newline
G.W.~Wilson, Ergod. Theor. \& Dyn. Sist. {\bf 1} (1981) 361; \newline
M.F.~de Groot, T.J.~Hollowood and J.L. Miramontes,
Commun. Math. Phys. {\bf 145} (1992) 57.}
\Rref{MASS}{C.R.~Fern\'andez-Pousa, M.V.~Gallas, T.J.~Hollowood, and
J.L.~Miramontes, {\sl ``The Symmetric Space and Homogeneous Sine-Gordon
Theories''\/}, hep-th/9606032 to appear in Nucl. Phys.~{\bf B}.}
\Rref{WITTEN}{E.~Witten, Commun. Math. Phys. {\bf 92} (1984) 455.}
\Rref{BAIS1}{F.A.~Bais, P.~Bouwknegt, M.~Surridge, and K.~Schoutens,
Nucl. Phys. {\bf B 304} (1988) 348.}
\Rref{BAIS2}{F.A.~Bais, P.~Bouwknegt, M.~Surridge, and K.~Schoutens,
Nucl. Phys. {\bf B 304} (1988) 371.}
\Rref{KZ}{V.G.~Knizhnik and A.B.~Zamolodchikov, Nucl. Phys. {\bf B 247}
(1984) 83.}
\Rref{CARDY}{J.L.~Cardy, {\sl ``Conformal invariance and statistical
mechanics''\/}, in {\it ``Fields, Strings and Critical Phenomena''\/},
eds. E.~Br\'ezin and J.~Zinn-Justin, Les Houches 1988, Session XLIX,
North-Holland, Amsterdam (1990) 169-245.}
\Rref{ZAM}{A.B.~Zamolodchikov, Adv. Stud. Pure Math. {\bf 19} (1989)
641; Int. J. Mod. Phys. {\bf A3} (1988) 743; JETP Lett. {\bf 46}
(1987) 160.}
\endref

\ciao